\documentclass[a4paper,12pt]{article}
\usepackage[margin=2.5cm,nohead]{geometry}
\usepackage[utf8]{inputenc}
\usepackage{natbib}
\usepackage{color}
\usepackage{sgame}
\usepackage{pdflscape}
\usepackage{graphicx}
\usepackage{amssymb}
\usepackage{amsbsy}
\usepackage{amsmath,amsthm}
\usepackage{booktabs}
\usepackage{rotating}
\usepackage{colortbl}
\usepackage{caption}
\usepackage{setspace}
\usepackage{wasysym}
\usepackage{longtable}

\usepackage{subcaption}
\usepackage{float}
\usepackage{pdfpages}
\usepackage{blindtext}
\usepackage{xcolor}
\usepackage{soul}
\usepackage{adjustbox}
\usepackage{threeparttable}
\usepackage{threeparttablex}
\usepackage{tabularx}
\usepackage{booktabs} 
\usepackage{makecell} 
\usepackage{colortbl} 
\usepackage{hyperref}
\usepackage{adjustbox}
  {\end{tabular}\par\medskip}

\begin{document}
\sloppy
\title{Friends, Key Players and the Adoption and Use of Experience Goods
}

\author{
    Rhys Murrian\footnote{Department of Economics and SoDa Labs, Monash University (\href{mailto:rhys.murrian@monash.edu}{rhys.murrian@monash.edu})} \hspace{5mm}
    Paul A. Raschky\footnote{Department of Economics and SoDa Labs, Monash University (\href{https://praschky.GitHub.io/}{praschky.github.io} and \href{mailto:paul.raschky@monash.edu}{paul.raschky@monash.edu})} \hspace{5mm}
     Klaus Ackermann\footnote{Department of Econometrics and Business Statistics and SoDa Labs, Monash University (\href{https://sites.google.com/site/acck1aus/}{sites.google.com/site/acck1aus/} and \href{mailto:klaus.ackermann@monash.edu }{klaus.ackermann@monash.edu })} 
    }

\date{September, 2024}
\maketitle

 
\begin{abstract}

\noindent This paper empirically investigates how an individual’s network influences their purchase and subsequent use of experience goods. Utilising data on the network and game-ownership of over 108 million users from the world’s largest video game platform, we analyse whether a user’s friendship network influences their decision to purchase single-player video games. Our identification strategy uses an instrumental variable (IV) approach that employs the temporal lag of purchasing decisions from second degree friends. We find strong peer effects in the individual game adoption in the contemporary week. The effect is stronger if the friend who purchased the game is an old friend compared to a key player in the friendship network. Comparing the results to adoption decisions for a major label game, we find peer effects of a similar size and duration. However, the time subsequently spent playing the games is higher for players who were neither influenced by a peer who is a key player nor an old friend. Considering the increasing importance of online networks on consumption decisions, our findings offer some first insights on the heterogeneity of peer effects between old and key player friends and also provide evidence in consumers' biases in social learning.


\medskip 

\noindent \emph{Keywords:} networks, experience goods, product adoption, taste projection  \medskip 

\noindent \emph{JEL classification: D12, Z13}

\end{abstract}

\pagebreak

\onehalfspacing

\captionsetup[figure]{labelfont=bf} 
\captionsetup[table]{labelfont=bf} 

\section{Introduction} \label{intro_1}
Information about an experience good’s quality is often difficult to infer before the consumer has used or purchased such a good. This creates a situation of information asymmetry between the buyer and the seller which can have adverse effects on the efficiency of the market. In such cases, consumers often rely on information from friends who have previously purchased the experience good and can share information about the quality of the good and lead to peer effects in purchase decisions \citep[e.g.][]{campbell2020word}. However, individuals often overestimate the degree to which others share their preferences, leading to biased inferences about product quality in market settings where social learning is prevalent. Consumers might overestimate how much they will enjoy an experience based on their peers' purchases which can lead to distortions in the market \citep[e.g.]{gagnonbartsch2024}.

The purpose of this paper is bridge and build an empirical complement to these two phenomena. In particular, we use data from the world’s largest digital gaming platform, Steam, that contains information about the online friendship network and game-ownership of over 108 million users of the Steam platform. We combine this data with additional scraped data on users’ game usage to construct a balanced panel dataset at the individual and week level for 269 weeks over the period September 2008 to October 2013. Using a combination of fixed effects and an instrumental variable (IV) to instrument for friends\' purchasing outcomes, we estimate a linear-probability model that identifies the likelihood of a user purchasing a particular game, dependent upon whether the user’s online friends purchased that game in the current or previous weeks. Our results show that a user's purchase of a particular game in a given week is positively impacted upon by the purchasing decision of the user’s friends in the current and previous weeks.  Comparing the results to adoption decisions for a major label game, we find peer effects of a similar size and duration. The effect is stronger if the friend who purchased the game is an old friend compared to a key player in the friendship network. Using this information, we then construct a cross-sectional dataset and analyse the time subsequently spent playing the game.


The video game industry itself has seen an explosion in growth over the last decade, and with the COVID-19 pandemic, is only expected to increase in size.  In 2020, the video game industry was valued at \$173.7 billion US dollars and is forecasted to grow to \$314.4 billion US dollars by 2026, an average annual growth rate of 9.6\% \citep{mordorintelligence}. For comparison, in 2018, global box office revenue from Hollywood-based films was approximately \$41.7 billion US dollars whereas global revenue generated by the video game industry was approximately \$151.2 billion US dollars \citep{dautovic_2021}. Further, it is estimated that there were over 2.5 billion people playing some form of video game in 2020 \citep{dautovic_2021}, over a quarter of the global population. As such, the importance of the video game industry to the global economy must not be understated, especially in relation to the entertainment sector.

While there are multiple platforms across the video game industry, the most dominant platform is Steam, which is hosted by the Valve Corporation.  Steam has dominated the PC gaming market for a number of years, with approximately 47\% of video game publishers selling their titles on Steam \citep{gdc2019}.  Given its dominance in the video game market, the number of Steam users has risen exponentially since its inception in 2003, with over 120 million active users in 2020 \citep{steam2021review}.  Users on Steam may enhance their gaming experience by becoming friends with others that they meet online.  Thus, Steam not only serves as a video game distribution platform, but also as an online network.

Steam provides an opportunity for developers, publishers and researchers alike to utilise its data.  Data related to different users and apps can be obtained via the Steam Web API where users have a public profile \citep{valvedev_steamwebapi_page}.  Examples of such data include a summary of a user’s profile, the games they own, their playtime on Steam over the last two weeks, amongst others.  The friends of a user on Steam are also included as downloadable data.  As such, a network of public users across the Steam platform can be compiled to construct an online network for the Steam platform.

The video game market on Steam provides an opportunity to empirically explore peer effects in a market for experience goods.  Video games fall under this category as utility derived from a video game is consumed once a user has played the game.  As such, the quality of a video game is not fully observable before a user purchases it\footnote{This is not always the case, as information may be available to users before they purchase a game via reviews from established gaming websites. However, for games that are not as well-known, such as games under the indie genre, these sources may provide little information.}.  In such instances, word-of-mouth recommendations through social networks become a vital source of information for an individual \citep{campbell2020word}.

Our analysis focuses on a distinct genre of games known as \emph{indie games}. Indie games cover a genre of games produced by small-scale developers with limited financial support. As such, indie games generally have little-to-no advertising, making it difficult to compete with triple-A or blockbuster titles, such as the \emph{Call of Duty} series. However, word-of-mouth has been an avenue for indie games to increase their popularity, allowing them to compete with blockbuster titles \citep{mcelroy_polygon_2013}. Thus, peer effects are likely to be prevalent for purchasing decisions for games in the indie genre. We focus our analysis on one indie game in particular, a single-player platformer called \emph{Super Meat Boy} (SMB).\footnote{While Steam offers a number of multiplayer indie games (e.g. \emph{Terraria}), we do not analyse how peer effects impact adoption and playtime decisions for these game types due to the inherent likelihood of the existence of peer effects in cases where experience goods are jointly consumed.} SMB was released in November 2010 by the developer and publisher \emph{Team Meat}. SMB eventually became a well-known indie game, selling over one million copies by January 2012.\footnote{\textit{``Fun Fact: Super Meat Boy past the million sales mark last month!''}, 2012-01-03,https://x.com/SuperMeatBoy/status/154091784929161217} The popularity of SMB provides a useful case study to investigate whether peer effects played a significant role in influencing users’ purchasing behaviour. As a point of comparison, we estimate the peer effects of adoption and usage for \emph{Fallout: New Vegas} (NV), a single-player game released in October 2010 by the triple-A publisher, Bethesda Game Studios.

The outline for the rest of this paper is as follows: Section 2 presents a literature review on studies related to our paper, Section 3 provides background information on the Steam platform and network, Section 4 presents a showcase of the data involved with the analysis, Section 5 details the empirical strategy undertaken, Section 6 showcases the results, and Section 7 concludes.

\section{Related Literature}  \label{related_lit_1}
 
Our paper adds to a large body of literature on peer effects within economics. Peer effects have been studied across a wide range of topics, including education \citep{calvo2009peer,patacchini2017heterog,hanushek2003does,burke2013classroom}, labour market outcomes \citep{munshi2003networks, mas2009peers}, and consumption \citep{de2020consumption, agarwal2021thy, roth2015conspicuous}, amongst others. In particular, our paper relates to peer effects on product adoption. Previous studies have primarily focused on product adoption with regard to technology in developing economies, such as agricultural innovation \citep{foster1995learning,conley2010learning,barrett2016system}, cookstoves \citep{miller2015learning}, and microfinance services \citep{banerjee2013diffusion}. Technology adoption has also been studied in developed economies, especially in the case of green energy sources \citep{bollinger2012peer,bollinger2022visibility, lanauze2021}. 

A limited number of studies have examined peer effects on product adoption with experience goods. \citet{bailey2022peer} use friendship data from Facebook to estimate peer effects on phone purchases; \citet{ameri2019word} estimate peer effects on the adoption of anime series based off an anime fansite where users are able to become friends and review anime series; and \citet{moretti2011social} explores how peer effects impact upon watching movies at the theatres based off box office sales. Peer effects have even been estimated in the adoption of video games \citep{li2022peer,tudon2022distilling}. Both of these studies use data from the Steam platform to estimate peer effects, however they differ in multiple ways from our paper. Firstly, \citep{tudon2022distilling} outlines a method to estimate peer effects through a discrete-choice model where only aggregated relational data \citep[see][]{breza2020using} is required. Secondly, \citet{li2022peer} paper is part of a large body of peer effects literature which falls under a non-economic stream. The authors employ a framework that differs from the conventional linear-in-means model for peer effect estimation within the economic literature and use cross-sectional network and games purchased data in their method. In contrast, our paper employs a conventional linear-in-means model and uses panel data for both the network and games purchased data at the weekly level to estimate both contemporaneous and lagged peer effects of video game adoption.

Network structure is another dimension along which peer effects have been investigated. Heterogeneous peer effects have been examined for various network properties, including centrality \citep{banerjee2013diffusion,katona2011network,miller2015learning, robalino2018peer,zenouislam_repec} and tie quality or strength \citep{patacchini2008strength,patacchini2017heterog, siciliano2016s}. A small number of of studies have considered heterogeneous peer effects for both centrality and tie quality simultaneously. In an unpublished study, \citet{hoffman2017} examines the uptake of medical check-ups within disadvantaged neighbourhoods in the Philippines. The author uses a discrete-time hazard model to estimate heterogeneous peer effects in regard to both tie strength and centrality. Findings from the research indicate that was no heterogeneity in regard to tie strength, though, there was heterogeneity in regard to centrality, whereby more central peers were estimated to have a stronger influence on an individual going for a medical check-up compared to non-central peers. Outside of the economic literature, \citet{shakya2015social} examine peer effects within the context of latrine ownership across a network of rural villages in India. The authors find that stronger ties with peers who own a latrine is correlated with an individual owning a latrine. Further findings from the study indicate that the centrality of individuals affects how much they are influenced by their peers. 

To the best of our knowledge, our paper is the first to consider how peer effects impact upon both the adoption decision and subsequent experienced utility of an individual. This presents a major contribution to the literature as previous studies have focused on peer effects on product adoption decisions without measuring the utility experienced from these decisions. As such, we highlight a distinction between expected utility from product adoption due to peer influence in contrast to actual utility realised from experiencing the good. Further, we examine this distinction heterogeneously, comparing adoption decisions and utility experienced under the contexts of no peer effects, peer effects induced by key players, and peer effects induced by old friends.





Our analysis of the heterogeneous effects of peer's influences on the subsequent use (playtime) from the purchase also offers a complementary perspective to the literature analysing how individual's often misjudge the degree to which others’ tastes are similar to their own \citep[e.g.,][]{ross1977false, engelmann2012deconstruction,orhun2013conditional,gagnonbartsch2024} In particular, we relate to the recent work by \citet{gagnonbartsch2024} who analyse how consumers' biases in social learning (``taste projection'') affect their perception of a product's quality and their subsequent purchase decision. Individuals project their own tastes onto others, leading to misperceptions about how widely shared their preferences are. This projection bias results in distorted beliefs about a product's quality based on observed market behavior, such as the number of other consumers purchasing the product at a certain price. Their findings suggest that this bias leads to a self-fulfilling cycle where individuals' inferences about product quality are shaped by their idiosyncratic tastes, which in turn influence their purchasing decisions and market demand in a way that may not accurately reflect the true value of the product. Individuals often overestimate the degree to which others share their preferences, leading to biased inferences about product quality in market settings where social learning is prevalent. Our findings support the propositions by \citet{gagnonbartsch2024}, showing that the subsequent engagement with the product is lower if the initial purchase was driven by peers. Users who did not have any peers purchasing the game, showed a systematically higher engagement (play time) with the game.


Our paper also presents a contribution to the literature on peer effects identification. Identifying causal impacts of peer effects has long been studied within the economic literature due to the number of challenges that need to be overcome in any causal estimation of peer effects. One of the key challenges outlined by \citet{manski1993identification} is the 'reflection problem', which many previous studies have resolved through the use of an IV. \citet{bramoulle2009identification} showcase that where individual-level network data is available, the characteristics of second-degree friends can be used as an IV to resolve the reflection problem. Their finding presents a generic use case, and has been employed in subsequent studies \citep[e.g.,][]{de2010identification, patacchini2017heterog,hoffman2017}. We propose a similar IV to \citet{bramoulle2009identification} to resolve the reflection problem that is only applicable where individual-level panel network data is available. Peer effects under a panel network context has received little attention thus far in the literature, with the exceptions of \citet{comola2021treatment}, who propose a model to identify and estimate causal peer effects where a treatment influences change in the structure of the network, and \citet{depaula2024_panelsocialties}, who propose a model to identify social ties from panel data observations (where no social network data is available) and apply it empirically to state-based tax competition within the US. Further, \citet{bramoulle2020} provide a brief review of the peer effects literature under the context of panel data.

Finally, our results also add to a relatively small literature on computer games in economics.  \citet{aguiar2021leisure} built a theoretical, labour supply model that shows how the availability of video games (a relatively cheap leisure activity) contributes to a downward shift in labor supply among young males in the US.  Other studies within the economics literature have mainly focused on the relationship between video gaming and violent behaviour or crime \citep[e.g.,][]{cunningham2016violent,ward2011video}.\footnote{In addition, there is a very large literature in psychology on the mental health effects of video games. For example, see \citet{granic2014benefits} for a review of the literature).} We speak to these finding, by showing how peer effects lead to a wider adoption of video games. Our paper is also part of a growing number of studies in economics that harnesses information on individual behaviour and revealed preferences from digital footprint data. For example, \citet{Gandhi2025} investigate the impact of a player's evolving beliefs about winning on their engagement with a video game. Using a large dataset of \textit{League of Legends} matches, the authors track the changes in players' beliefs about their likelihood of winning throughout a match. They then examine how these belief trajectories, characterized by measures like suspense (expectation of an important event) and surprise (deviation from expectations), influence player engagement, measured by their decision to play another match. They find that both suspense and surprise contribute positively to player engagement, suggesting that games that generate fluctuating and unpredictable belief paths are likely to be more engaging.

\section{The Steam Platform and Network}  \label{steamplatform_network_1}
Launched in 2003, the Steam platform provides a digital marketplace for publishers to sell video games and for users to purchase, download, and play them. The platform is available on PC for Windows, MAC, and Linux systems, as well as mobile phones. Signing up for an account on Steam is free, however, purchasing a game typically requires a pecuniary cost to the user.\footnote{Some games on Steam are free-to-play and require no pecuniary cost to the user. An example of this is \emph{Team Fortress 2}.} Upon sign up, users must set a unique username and avatar for their Steam profile.\footnote{Setting an avatar is optional, but a unique username is required for all Steam users.} In 2020, there were over 120 million active users and 43,000 titles (games and game-related apps) on Steam, with 47\% of video game publishers selling their titles on Steam. 

As well as being a digital marketplace (see Figure \ref{appfig:steamstore}), Steam also operates as an online social network, where users can send friend requests and become friends with other users on the platform (see Figure \ref{appfig:steamfriends}). The friendship mechanism is similar to that of Facebook's, where a user sends a friend request to another user and the latter can then decide whether to accept or reject the friend request. If the friend request is accepted, a bidirectional friendship link is formed between the two users. A friendship limit of 250 friends is set by default on the platform.\footnote{The default friendship limit for a user can be extended to a maximum of 2.000 friends through levelling up their Steam profile. Further detail is provided in Appendix \ref{appendix_sect_data}.} Friendships are formed via in-game interactions with other users on Steam, as well as other means, such as through meeting other users via mutual friends.\footnote{An example of where this can occur is through the `Group Chat' function on Steam.} 

The Steam platform also provides a way to showcase a user's advancement through a game via achievements.\footnote{Not all games on Steam have a set of achievements for players to unlock. An example of this is \emph{Half-Life}.} Achievements are obtained by completing certain objectives within a game. Once an achievement is obtained, the achievement is displayed on a user's profile with the exact timestamp of when the achievement was unlocked. Generally, achievements start off as quite easy to obtain, but progressively become more difficult as a user progresses further in a game.

Using data from video game platforms such as Steam for economic research  has a number of advantages: First, they contain very granular individual-level data providing information about individual online networks and relatively high-frequency consumption data. Second, an increasingly large fraction of non-durable (and even durable) goods are bought online. Third, the data contains revealed information about an individual's social network online (as opposed to stated data from surveys). Fourth, video game platform data further records a wealth of other digital footprints that can be used to measure individual behaviour and personal traits (e.g. cheating).

\section{Data}  \label{data_1}

Our base dataset comes from \citet{ONeilletal2016}, who undertook multiple crawls of a number of Steam services between 2013 and 2014, including the Steam Web API, Steam Storefront and XML data from steamcommunity.com. The resulting dataset from these crawls was made publicly available in 2016.\footnote{Data available for download at https://steam.internet.byu.edu/ (as of March 16, 2022).} Table \ref{table:oneill_dataset_desc} in Appendix \ref{appendix_sect_additional_tables} provides a descriptive summary of the \citet{ONeilletal2016} dataset.

Four tables from the \citet{ONeilletal2016} dataset were used as part of the analysis: \emph{Player\_Summaries}, \emph{App\_ID\_Info}, \emph{Friends}, and \emph{Games\_2}. The Player\_Summaries table was used to obtain the universe of Steam users at the time of the first crawl undertaken by O'Neill et al. (2016) in February 2013. Along with each user's Steam ID, the Player\_Summaries table also provides attributes of each user, such as the name of their Steam profile, privacy status, and the time their Steam profile was created. The App\_ID\_Info table was used to obtain the Steam App IDs and release dates for SMB and NV. The Friends table was used to obtain the friendships between Steam users and the timestamp for when users became friends with each other. The Games\_2 table was used to obtain the different games owned by each Steam user in the dataset at October 2014. This data was used to identify all Steam users in the dataset that own SMB and NV, respectively.

The base dataset was combined with player achievement data from Steam, which was extracted via the Steam Web API. Player achievements within SMB and NV were extracted for all Steam users who own SMB and NV, where possible.\footnote{Achievement data can only be extracted for users who currently have a public Steam account.} To access the API, multiple API keys were generated using personal Steam accounts. Other parameters required to extract player achievements included each user's Steam ID and the Steam App IDs for SMB and NV.\footnote{See https://partner.steamgames.com/doc/webapi/ISteamUserStats for further detail.} Figure \ref{appfig:steamachieve} in Appendix \ref{appendix_sect_steamplatform} describes how achievement data for each game was scraped from public player's profiles. 

The original \citet{ONeilletal2016} data only contains information about the games that each player has purchased but not the time of the purchase. Instead, we use the date (based on the timestamp) of the player's first achievement in the respective game (SMB and NV) as an approximation for the purchase date of the game. It is fairly reasonable to assume that the date of the purchase and the date of the first achievement in the game are very close together. First, it is very common for players to immediately play a game upon purchasing and downloading it.\footnote{SMB also has relatively small disk space requirements, and as such, the time to complete the download is relatively short.} Second, the time required to unlock the first achievement is also relatively short. We provide evidence of this in Appendix \ref{appendix_sect_data}. Third, considering that we are aggregating the data up to the weekly level, it is even more reasonable to assume that the date of the first achievement is within the same week as the purchase date.

One concern with this approach is that we not only capture SMB/NV purchases but also events where a player received the game as a gift. Using statistics from `Steam Tracker'\footnote{ https://steam-tracker.com/app/40800/. This website tracks the game library and inventory of a sample of around \~ 20,000 steam users.} shows that for the games used in our analysis, gifting is relatively rare and only accounts for about 3.5\% of SMB ownership.

From the Games\_2 table, there were approximately 1.06 million Steam users who had purchased Super Meat Boy at the time \citet{ONeilletal2016} undertook their second crawl in 2014. Combining this with the player achievement data for Super Meat Boy for each of these users, our sample reduces to 166,560 Super Meat Boy players with a purchase date prior to October 2013.\footnote{Player achievement data for the majority of Super Meat Boy owners (approximately 75\%) from the base dataset was not publicly available due to Steam's private policy change in 2018 (see \citep{humphries_pcmag_2018}). The policy change sets a user's privacy setting to \emph{Friends Only} by default. As such, player achievements can only be scraped for users who opted out of the default privacy setting and made their profile public. The remaining Super Meat Boy players excluded from the sample did not obtain any achievements within Super Meat Boy or obtained their first achievement after October 2013 (approximately 8\%).} To reduce the size of the Steam user sample for the analysis, we randomly selected 44,773 players from the treatment group and 44,773 players from the control group. 


\subsection {Defining the Steam Network}

To identify friends, key players as well as construct our instrumental variable, we used the information provided \emph{Player\_Summaries} and \emph{Friends} tables to derive the nodes and edges for the Steam network used in our sample. From the \emph{Player\_Summaries} table, 108.8 million Steam users were existent within the base dataset, representing the universe of Steam users at the time O'Neill et al. (2016) undertook their initial crawl. Out of this total number of Steam users, 101.9 million had ‘public’ accounts, inferring that 93.6\% of users at the time of collation had publicly available data.\footnote{Data for the remaining 6.9 million 'private' accounts was unavailable, including games owned and friendships with other 'private' accounts. As such, the data collected for the Steam network is not fully complete, but it represents the majority of the network. Further detail is provided in the Appendix.}  We reduce the number of Steam users for our analysis further by excluding stagnant or bot accounts; that is, accounts without any playtime on Steam. This reduces our sample to 42.2 million Steam users, which we define as nodes in the network. 

From the \emph{Friends} table, 196.4 bidirectional friendship links were existent within the base dataset, representing the universe of friendship links on Steam. Similar to the number of Steam users, we reduce the number of friendship links for our analysis by excluding all private accounts and stagnant accounts. This reduces the network to 122.2 million friendship links, which we define as edges in the network. Out of the 42.2 million nodes in the network, 22.1 million have at least one friendship link. The remaining 20.1 million nodes in our sample were assumed to have no friends.

For the construction of our instrumental variable we required information on each user's second-degree friends. Given the size of the network, the second-degree friends of players in the treatment and control groups in each week were calculated using GPU parallel computing\footnote{The Python library \emph{RAPIDS cuGraph} was used for the second-degree friends calculation.} via a high-performance computing (HPC) cluster at Monash University.\footnote{See https://www.massive.org.au for further detail.} Player achievements for SMB were then merged to all second-degree friends who purchased SMB. The timestamp of the earliest achievement for SMB for each second-degree friend was then used to calculate the week in which each of the second-degree friends purchased SMB, where relevant.

For the heterogeneity analysis, the key player friend variable was constructed by identifying which Steam users in the network were key players four weeks prior to the release of SMB to maintain exogeneity. The Katz centrality measure for each Steam user was calculated as a first step. As a second step, we defined a binary variable for Steam users who had a Katz centrality measure within the 99th percentile or above. Thirdly, we interacted the key player binary variable with the friends of each of the 89,546 Steam users in our sample. If a Steam user in our sample had a friend who purchased Super Meat Boy first and that friend was a key player, then the Steam user would be considered as having a 'key player' treatment as part of the heterogeneity analysis.


The old friend variable was constructed by identifying which friends in a Steam user's account had been friends with the user for at least one year at the time of four weeks prior to the release of SMB in order to maintain exogeneity. We defined a binary variable for friends of each Steam user who had been friends with the Steam user for over a year. We then interacted the old friend binary variable with the friends of each of the 89,546 Steam users in our sample. If a Steam user in our sample had a friend who purchased Super Meat Boy first and that friend was an old friend, then the Steam user would be considered as having an 'old friend' treatment as part of the heterogeneity analysis.



\subsection {Summary Statistics} 

Table 1 presents summary statistics for the sample used in our analysis. The mean probability of a Steam ID purchasing Super Meat Boy within the panel period is approximately 2.17\%. Further, the mean probability of a Steam ID purchasing Super Meat Boy within a given week is approximately 0.01\%. The mean number of games purchased by a Steam ID is approximately 21 games, highlighting that Steam users are generally active on the platform. On average, a player within the sample has approximately 17 friends, with the maximum having 359 friends, which is above the standard 250 friendship limit. Out of the average 17 friends, approximately 5 will be friends who are key players in the network, while approximately 7 will be 'old' friends. Moreover, the mean probability of a player being treated by a key player friend purchasing Super Meat Boy is approximately 12.08\%, whereas the mean probability of being treated by an old friend purchasing Super Meat Boy is approximately 4.06\%, highlighting that Super Meat Boy is more often purchased by key players in the Steam network. 










\begin{table}[!h]
\centering
\caption{Summary Statistics}
\begin{tabular}{@{\extracolsep{5pt}}lcccccc}

\\[-1.8ex]\hline \hline \\[-1.8ex]
\textbf{Variable}	&	\textbf{Obs}	&	\textbf{Mean}	&	\textbf{Std. dev.}	&	\textbf{Min}	&	\textbf{Max}	\\\hline

SMB&	89,546	&	0.022	&	0.145	&	0	&	1	\\
NV&	89,546	&	0.019	&	0.136	&	0	&	1	\\
treatment	&	89,546	&	0.500	&	0.500	&	0	&	1	\\
\# of Games &	89,546	&	20.625	&	36.997	&	1	&	1011	\\
\# of Groups &	89,546	&	5.172	&	17.860	&	0	&	952	\\
Start week &	89,546		&	35.809	&	57.559	&	0	&	113	\\
Friends&	89,546	&	2.150	&	3.560	&	0	&	134	\\
KP Friends&	89,546	&	0.772	&	1.491	&	0	&	109	\\
Old Friends&	89,546	&	1.128	&	2.290	&	0	&	134	\\\hline
SMB purchase&	13,790,084	&	0.013	&	0.113 &	0	&	1	\\
NV	purchase&	13,790,084	&	0.011	&	0.106	&	0	&	1	\\

\hline \hline \\ 
\end{tabular}
\end{table}

\begin{figure}[!h]%
\centering 
\caption{SMB Purchases Over Time}
 \includegraphics[width=12cm]{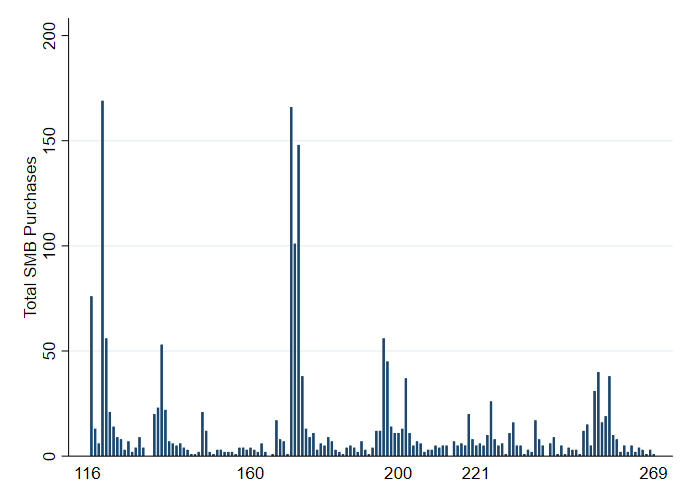} %
    \label{fig:smb_pur_time}%
\end{figure}

Figure \ref{fig:smb_pur_time} presents the weekly number of total SMB purchases over our sample period (weeks 116 -- 269). The spikes in purchases in the early weeks reflect large purchase numbers around a game's release, where games are normally promoted by the platform as well as through reviews in gaming magazines. The peaks in between weeks 160 and 180 are likely due to the release of SMB on the MAC OS which opened the product for an entire new market segment. It is important to note, that our specification includes week fixed effects which absorb these global, time specific shocks to individual purchase decisions.

\begin{figure}[!h]%
\centering 
\caption{Number of Friends}
 \includegraphics[width=12cm]{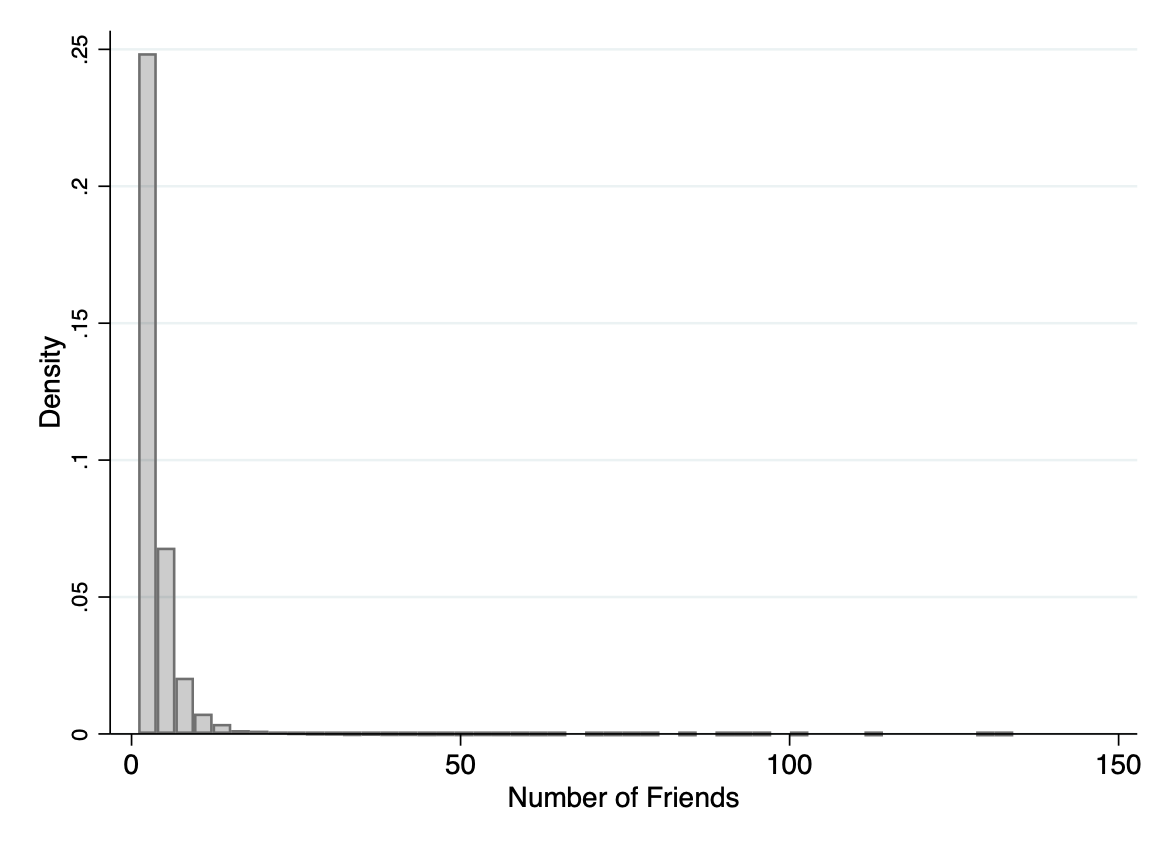} %
    \label{fig:histo_friends}%
\end{figure}

Figure \ref{fig:histo_friends} plots the distribution of the number of friends of each player in our sample \ref{fig:histo_friends}. The vast majority of steam users in our sample, and the total population of steam users, do not have any friendship connections on the steam platform. The average number of friends is around 2 with a maximum of 134 friendship connections.

 \pagebreak

\section{Empirical Framework}  \label{empiricalframework_1}

The starting point of our empirical framework is \citet{manski1993identification} linear-in-means model which is considered the workhorse model within the literature to estimate peer effects (\citep[e.g.,][]{blume2015linear,ryan2017measurement}). Given the panel structure of our data, we adopt the generic framework for network panels posited by \citet{bramoulle2020} and specify the following econometric model: 


 
\begin{equation}
y_{i,t} = \alpha_{i}  + \beta \sum_{t=0}^{n} g_{ij,t} y_{j,t} + \mathbb{W}_{t} + \epsilon_{i,t} 
\label{eq:1} 
\end{equation}


Here, $y_{i,t}$ is a binary dependent variable indicating whether an individual \emph{i} has purchased (adopted) SMB or not in week \emph{t} or not.  $\alpha_{i}$ is a vector of individual player fixed effects that controls for any unobserved, time-invariant characteristics of individual \emph{i} that might drive their likelihood to purchase the game. 
$g_{ij,t}$ is the network of individuals \emph{j} connected to individual \emph{i} in a given period \emph{t}. $\alpha_{j}$ is a vector of player fixed effects related to friends \emph{j} of individual \emph{i}. This controls for any unobserved, time-invariant characteristics of friends j that may influence player \emph{i} to purchase the game. $y_{j,t}$ is a binary independent variable indicating whether a friend \emph{j} of individual \emph{i} purchased SMB in week \emph{t} or not. In our preferred specification, $\mathbb{W}_{t}$ is a vector of time fixed effects that will consist of 269 (-1) year-week dummies. This set of time-fixed effects controls for any unobserved global shocks, (e.g. the week of the release of the game itself, publication of reviews, releases that are close substitutes to game etc.) that affect the entire network. $\beta$ is our parameter of interest that will be estimated. A statistically significant and positive $\beta$ indicates that the purchase decision of the player’s friends increases the purchase likelihood of player \emph{i}, otherwise known as an endogenous peer effect. Standard errors are clustered at the player level. The focus of the analysis is on the effect of friends' $j$ decision to adopt the small independent game, SMB, on player $i$'s adoption decision of SMB.  

Further, we define treatment and control groups as part of our estimation. We consider players who have at least one or more friends that purchased SMB as part of the treatment, while players who have friends, but none of them purchased SMB, are considered to be part of the control group. Given the scale of the Steam network, we randomly select 50,000 players who fit the treatment criteria to be part of the treatment group and randomly select 50,000 players who fit the control criteria to be part of the control group. As a result of data cleaning, the number of players drop to 44,773 within each group\footnote{5,227 players are removed from the original randomly selected 50,000 in each group due to the SMB purchase dates of their friends being later than October 2013, which is when the Steam network data stopped being collected.}.

\subsection{Challenges to identification} 
Existing literature on the estimation of peer effects \citep[e.g.,][]{bramoulle2009identification,bramoulle2020} have discussed two major challenges with appropriate estimation of peer effects: The reflection problem and correlated effects. Our identification strategy addresses both of these major challenges.



\subsubsection{Reflection problem}  \label{reflectionproblemsection}
First identified by \citet{manski1993identification}, the reflection problem refers to an identification issue in peer effect studies, whereby it cannot be determined if the outcome of an individual influences the outcome of their peers, or vice-versa, thereby leaving peer effects unidentified under a causal framework. Under a cross-sectional setting, the reflection problem is resolved under a network structure where individuals each have their own reference group. In such a case, second-degree friends' (or third-degree friends if network fixed effects are required) characteristics can be used to instrument for first-degree friends' (i.e. peers) outcomes \citep{bramoulle2009identification}. However, under a panel setting, the reflection problem is relevant for the contemporaneous period only. Peer outcomes in previous periods can readily be estimated on an individual's outcome in the current period, given the known direction of causality. That is, past peer outcomes can influence an individual's current outcome, but the opposite cannot hold true.


Similar to the cross-sectional setting, a network-based instrumental variable (IV) approach can be undertaken to address the reflection problem within the contemporaneous period. Two criteria must be fulfilled for instrument $z$ to be valid: $z$ has to be correlated with $y_{j,t}$ (relevance assumption); and $z$ has to be uncorrelated with $\epsilon_{i,t}$ (exclusion restriction). While the IV proposed by \citet{bramoulle2009identification} can be utilised in our setting,  we propose another instrument to address the reflection problem, which is only feasible under a panel setting. Our proposed IV utilises the network from the previous period and uses the SMB purchasing decision outcome from second-degree friends as an instrument. This is notionally defined as $y_{k,t-1}$, where k is a second-degree friend of i. The instrument $y_{k,t-1}$ fulfils both criteria for a valid instrument. Firstly, the purchasing decision of k in the previous week would directly affect j’s purchasing decision in the current week. Given that k's purchase occurred in the previous week, the reflection problem is non-existent. Secondly, the purchasing decision of k in the previous week cannot have a direct impact on i’s purchasing decision, as i does not observe k’s purchasing decisions, which satisfies the exclusion restriction. While our IV is derived through the network structure, similar to the IV proposed by Bramoull\'e et al. \citet{bramoulle2009identification}, we utilise previous period outcomes from second-degree friends as compared to second-degree friend characteristics from the contemporaneous period. Our proposed IV is more applicable under a panel setting with network data and limited information on individual characteristics.

\subsubsection{Correlated effects} 
There are two types of correlated effects that impact upon identification in peer effect studies: Common shocks and endogenous peer selection \citep{bramoulle2020}. Common shocks refer to exogenous shocks that impact all individuals across the entire network, leading to a change in individuals' behaviour. If unaccounted for, this change of behaviour resulting from a common shock, will confound peer effect estimates as it will be unknown as to whether the common shock or the peer effects led to a change in an individual's outcome. To address the issue of common shocks, we apply time fixed effects to absorb common shocks at each time period. Endogenous peer selection refers to the self-selection of peers by an individual who form said individual's reference group. Endogeneity arises from self-selection due to unobserved homophily, the tendency to select peers who share similar characteristics \citep{mcpherson2001birds,jochmans2023peer}. If unobserved shared characteristics between an individual and her peers are correlated with individual and peer outcomes, bias arises within peer effect estimates. To address the issues of endogenous peer selection, we apply individual fixed effects \citep{nair2010asymmetric,ma2015latent,bramoulle2020}. 


\subsection{Instrumental Variables Approach}

As detailed in section \ref{reflectionproblemsection}, the purchasing decision of friends in the contemporaneous period (\emph{$y_{j,t}$}) cannot be included as part of the identification due to the reflection problem. As such, we use the purchasing decision of second degree friends in the previous period (\emph{$y_{k,t-1}$}) as an IV for \emph{$y_{j,t}$}. We specify the following first stage:


\begin{equation}
y_{j,t} = \alpha_{i}   + \delta' \sum_{t=0}^{n} g_{jk,t} y_{k,t-1}  + \mathbb{W}_{t} + u_{i,t} 
\label{eq:2}
\end{equation}


The primary concern is to ensure that our IV approach fulfills the necessary assumptions of strength and validity to provide unbiased and consistent estimates.

\emph{Instrument Strength:} The strength of our instrument is supported by the network structure where the purchase decision of player $j$ influences the purchase decision of player $i$, and similarly, the decision of player $k$ influences the decision of player $j$. This cascading effect implies that player $k$'s decision, though indirectly connected, exerts a significant influence on player $i$'s purchase decision. This indirect influence is critical, as it strengthens our instrument by creating a meaningful variation in the explanatory variable that is correlated with the instrument.

\emph{Instrument Validity:} To ensure the validity of the instrument, we focus on second-degree friends, $k$, who are not directly connected to player $i$ but are only connected through player $j$. This separation is crucial because it helps to mitigate concerns that the instrument might be correlated with unobserved factors directly affecting player $i$'s decision.  The underlying assumption is that the primary connections in friendship networks, such as those on platforms like Steam, often originate from real-life relationships or other forms of direct interaction. By restricting our instrument to second-degree friends who are not directly connected, we reduce the risk that our instrument is endogenous to the same shocks or preferences that directly influence player $i$'s decisions. More importantly, we only use player $k'$s purchase decision in time $t-1$ which 

Furthermore, we acknowledge that players $i$ and $k$ might reside in the same geographic area, which could expose them to similar localized shocks, such as economic changes or region-specific events, that might influence game consumption. However, since Steam ads, releases, and bundles are globally distributed, these local shocks are less likely to align perfectly with the timing of game purchases.

\emph{Potential Shortcomings:} Despite the theoretical arguments above, there are caveats to this approach. The IV estimates only allows us to estimate a Local Average Treatment Effect (LATE), which means that they apply only to the subset of players whose purchase decisions are influenced by their second-degree friends. This subset is rare, as it consists of players who happen to have second-degree friends who also purchase the game. Consequently, the results may not be generalizable to the broader population of players. Additionally, there remains the possibility of unobserved, very localized, short-term shocks that might increase game consumption, potentially confounding our results.

\section{Results}  \label{results_1}
 
\subsection {Baseline Results} 
The estimates shown in Table \ref{table:baseline_results_1} highlight the baseline results. Both the first stage and second stage IV estimates, which is our preferred specification, are presented. The first stage estimate, which is shown in column 3, has an Anderson-Rubin test statistic of 341.17, signifying that the temporal lag of second degree friend $k$ adopting SMB is a very strong instrument for friend $j$ adopting SMB. The second stage estimate, which is shown in column 4, indicates that if a friend $j$ adopts SMB, there is a statistically significant increase of 1.7\% that player $i$ will adopt SMB in the same week. The results from the IV specification are compared to OLS and reduced form estimates, shown in columns 1 and 2, respectively. The reduced form estimate, which directly estimates the impact of the instrument on the outcome variable, showcases that the temporal lag of second degree friend $k$ adopting SMB has a statistically significant increase of 0.8\% on the likelihood of player $i$ adopting SMB in the current week. This result is not too dissimilar from the main IV results. However, the OLS estimate, which directly estimates the endogenous regressor on the outcome variable, showcases that friend $j$ adopting SMB results in a statistically significant increase of 7.3\% in player $i$'s probability of adopting SMB in the same week. This differs substantially from the main IV results. The discrepancy is likely a result of bias in the OLS estimates arising due to the endogeneity of the regressor.

\begin{table}[htbp]

\caption{Baseline Regression Estimates} \label{table:baseline_results_1}
\begin{adjustbox}{width=\linewidth, keepaspectratio}
{
\setlength{\tabcolsep}{10pt} 
\def\sym#1{\ifmmode^{#1}\else\(^{#1}\)\fi}
\begin{tabular}{l*{8}{c}}
\toprule
\toprule
 Dependent Variable:       & OLS & Reduced Form & IV First Stage & IV Second Stage \\
     & Adopt SMB$_{i,t}$ & Adopt SMB$_{i,t}$ & Friends j adopt SMB$_{i,t}$& Adopt SMB$_{i,t}$ \\
        & (1) & (2) & (3) & (4)  \\
\midrule
Friends $j$ adopt SMB$_{i,t}$ & 0.0733  &    &    & 0.0166  \\
   & (0.0008) &    &     &   (0.0009) \\
Friends $k$ adopt SMB$_{i,t-1}$  &   &   0.0077  & 0.0404 &  \\
 &  & (0.0013) &(0.0013)  & \\
\hline
Observations        &   13,700,538  &  13,700,538  &  13,700,538  & 13,700,538      \\
Player FE        &      \checkmark         &      \checkmark         &      \checkmark         &       \checkmark     \\
Week FE        &       \checkmark        &      \checkmark         &      \checkmark         &       \checkmark       \\
Anderson-Rubin stat.        &              &             &      341.17         &     \\
\bottomrule
\bottomrule
\end{tabular}
}
\end{adjustbox}

\vspace{6pt}
\begin{minipage}{\linewidth}
\footnotesize\textit{Notes:} The dependent variable is a binary variable indicating if player $i$ purchased SMB (except in the case of the first stage IV regression, where the dependent variable is a binary variable indicating if any of her friends $j$ purchased SMB). Clustered standard errors at a player level are in parentheses. All estimates are statistically significant at the .01 level.
\end{minipage}

\end{table}

\begin{figure}[!h]%
\centering 
\caption{Peer Effects by Type of Game}
    \includegraphics[width=12cm]{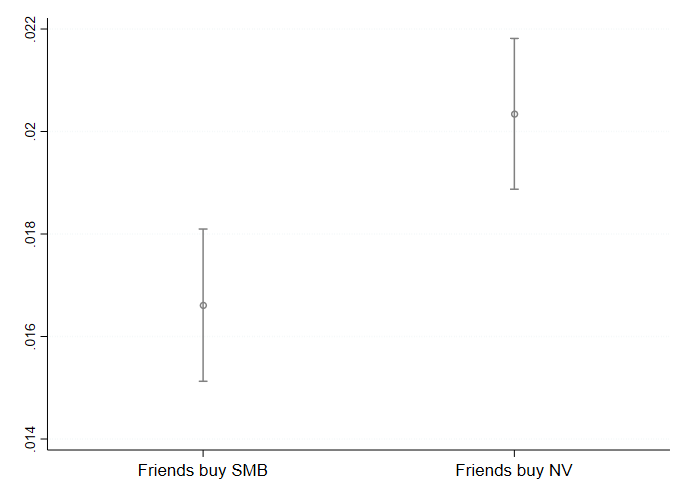} %
    \label{fig:base_smb_nv}%
\end{figure}

In Figure \ref{fig:base_smb_nv} we plot the estimated coefficients (IV) of the contemporary peer effects on purchase decisions for SMB as well as NV. If a friend purchases SMB, the likelihood of own SMB adoption increases from 1.3\% to around 3\% in the same week, reflecting the results from column 4 in Table \ref{table:baseline_results_1}. In comparison, if a friend purchases NV (large, blockbuster game), the likelihood of own NV adoption in a week increases from 1.1\% to 3.4\%. Interestingly, these results reveal that the extent of peer effects does not differ between the type of developer.

\pagebreak

\subsection {Heterogeneity by Type of Peers} 

We now conduct a heterogeneity analysis of the differences of peer effects by type of friends: We distinguish between \emph{Old Friends} and \emph{Key Player} friends. Considering that different types of friends have different knowledge of player $i$'s preferences as well as the quality and attributes of the game compared to other games, we hypothesise a systematic difference in peer effects on game adoption between Old and Key player friends. 

We extend the specification from (\ref{eq:1}) to the heterogeneity analysis regarding key players and old friends. We incorporate interaction terms for the key player dummy variable ($KP$) and treatment dummy variable as well as for the old friend dummy variable ($OF$) and treatment dummy variable. The model specification is as follows: 

\begin{equation}
y_{i,t} = \alpha_{i}   + \beta_1 \sum_{t=0}^{n} g_{ij,t} y_{j,t} \times KP + \beta_2 \sum_{t=0}^{n} g_{ij,t} y_{j,t} \times OF + \mathbb{W}_{t} + \epsilon_{i,t} \label{eq:3}
\end{equation}


Note that by including interaction terms with $KP$ and $OF$ simultaneously in equation \ref{eq:3}, the reference group are new (more recent) friends who are not key players in player $i$'s steam network.

\begin{figure}[htp!]%
\centering 
\caption{Old Friend vs. Key Player Friend SMB}
    \includegraphics[width=12cm]{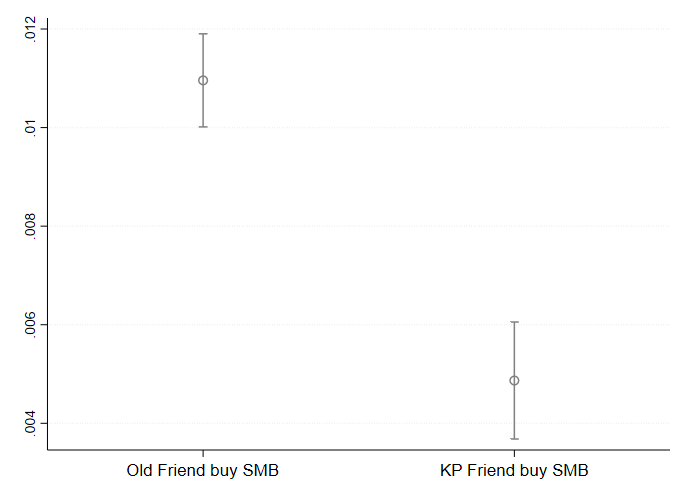} %
    \label{fig:het_oldkp_smb}%
\end{figure}

Figure \ref{fig:het_oldkp_smb} presents the results for SMB and shows that the peer effects are about twice as large compared to a key player friend. In other words, if player $j$ is an old friend of player $i$, she is twice as likely to purchase SMB compared to  player $j$ being a key player. The estimated coefficients are around 0.11 for old friends and 0.05 for key player, which are both smaller than the aggregate effect of around 0.13. Recall that the reference group in this setting are more recent friends who are not key players. This suggests that the strongest peer effects are actually from more recent friends.



\begin{figure}[htp!]%
\centering 
\caption{ Old Friend vs. Key Player Friend NV}
    \includegraphics[width=12cm]{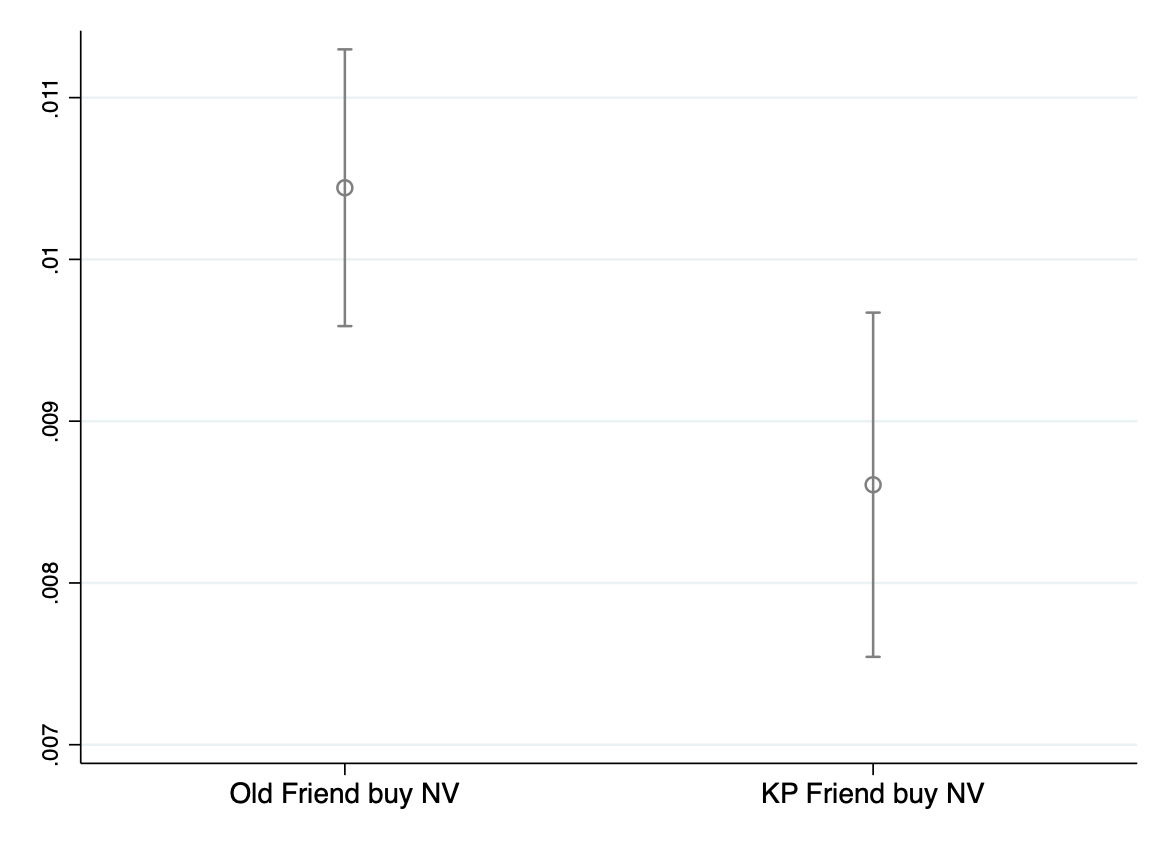} %
    \label{fig:het_oldkp_nv}%
\end{figure}


\pagebreak
\subsection {Peer Effects and Playtime}\label{sec:playtime} 

In the final section, we investigate how different purchases initiated by different type of peers lead to differences in satisfaction with the experience good. We use the information of total playtime for each game ($SMB$ and $NV$) from the steam database\footnote{This data is only available in the cross-section and measures the total playtime for each game at the time of the initial database scrape.}. To analyse the differential impacts of peer effects from either Key Players or Old friend on time subsequently played ($Playtime$), we specify the following, cross sectional, econometric model:

\begin{equation}
PT_{i} = \mu + \gamma_1 \left( Purchase \times KP \right)_i + \gamma_2 \left( Purchase \times OF \right)_i+ \mathbb{X}_{i} + u_{i}, 
\label{eq:pt} 
\end{equation}

where $PT_{i}$ is the log of hours that player $i$ spend playing the game. $( Purchase \times KP )$ and $( Purchase \times OF) $denote if the initial friend/peer who purchased the game was either a Key Player ($KP$) or and Old Friend ($OF$), respectively. $ \mathbb{X}_{i}$ is a vector of player $i$ specific covariates, including player $i$'s the number of games owned, number of groups, age of the steam account, no. of friends and if the player owns SMB or NV. The descriptive statistics of the data are presented in Table \ref{table:playtimedes} in the Appendix.

The three panels in Figure \ref{fig:playtime}  plot the estimated coefficients and 95\% confidence intervals from estimated equation \ref{eq:pt} for a combined sample of NV and SMB purchases (panel a), SMB only (b) and NV only (c). All three samples reveal the same pattern. Players who had a peer, Key Player or Old Friend, who purchased the game (either SMB or NV) before, and potentially were exposed to peer effects in consumption, do not depict a subsequent, systematically longer engagement (play time) with the game. In contrast, players whose purchase decision was not influenced by any peers (\textit{No Friend Purchase}) show a systematically longer engagement with the game. Relative to their counterparts whose initial purchase was influenced by peer effects, users without peer purchases show an approximately 45\% higher playtime in the case of SMB and 56\% higher playtime in the case of NV. On average, this translates to about 14 hours of more engagement with those games.  Note that the reference group in these specifications are players who had more recent friends, which are neither key players nor old friends, purchasing the game. In Table \ref{tab:pt}, we present results of including each of the peer effect variables individually and the estimated coefficient for the \textit{No Friend Purchase} variable remains qualitatively and quantitatively the same throughout all specifications.

While not a direct test of \citet{gagnonbartsch2024}, we think that the pattern of higher engagement by players who were not influenced by peer effects in consumption at least partially aligns the implications of their model. The concept of taste projections implies  that individuals often overestimate the extent to which others share their preferences, leading them to draw inaccurate conclusions about a product's quality based on its popularity. This ``taste projection'' is particularly impactful in markets for experience goods where the quality is difficult to assess before consumption. In our context of, users influenced by Key Players, who hold a more prominent position within the network, might be more susceptible to taste projection.
Users might also assume a greater alignment of preferences with Key Players or Old Friends, leading them to overestimate their own enjoyment of the game simply because a Key Player purchased it. This could explain the initial surge in playtime observed among this group. However, this initial enthusiasm driven by projected enjoyment might be short-lived. When the player's actual experience with the game fails to match their inflated expectations, which were based on a misjudgment of preference alignment, their engagement wanes, leading to a faster decline in playtime compared to those whose initial purchase decision was not driven by such projections. Overall, our results underscore the importance of considering taste projection and its influence on social learning when studying the adoption and consumption of experience goods within social networks.

There are a number of alternative explanations and underlying mechanisms that could explain these results. For example, the observed differences in playtime for SMB and NV could be influenced by game-specific mechanics that promote or hinder long-term engagement. For instance, an active online community might foster continued engagement through social interaction and competition, irrespective of the initial influence leading to the purchase. However, these other network effects should work against our results and lead to ultimately more engagement by users with peers who also purchased the same games. In their study using data on engagement with the online multiplayer game \textit{League of Legends}, \citet{Gandhi2025} show that players are more likely to abandon games when their perceived chance of winning is low. It is plausible that players influenced by Key Players or Old Friends, particularly in games like SMB and NV that are known for their difficulty, might experience a steep decline in engagement due to a mismatch between their skill level and the game's demands. This skill mismatch could lead to repeated negative experiences, lowering their perceived chance of winning and ultimately causing them to abandon the game sooner. There could also be other, unobserved factors that drive novelty decay and shifting preferences and correlate with a player's peers' decision to adopt a game. Players, regardless of their social influences, might be initially drawn to new games, dedicating significant time to exploring the new experience. As the novelty wears off, their engagement naturally declines.  If these unobserved factors are not captured by our set of control variables and they systematically correlate with different peer effect variables, this could be a potential source of a bias for our estimates.

 \begin{figure}
 \caption{Peer Effects and Playtime}
  \label{fig:playtime}
\begin{subfigure}[b]{0.45\textwidth}
    \includegraphics[width=7.5cm,height=27cm,keepaspectratio]{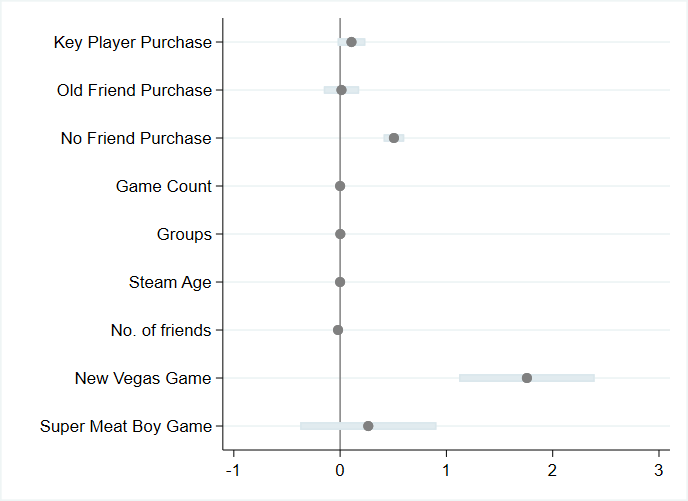}
    \caption{Playtime \& Peer Effects SMB and NV combined}
\end{subfigure}
	\qquad
	\begin{subfigure}[b]{0.45\textwidth}
    \includegraphics[width=7.5cm,height=27cm,keepaspectratio]{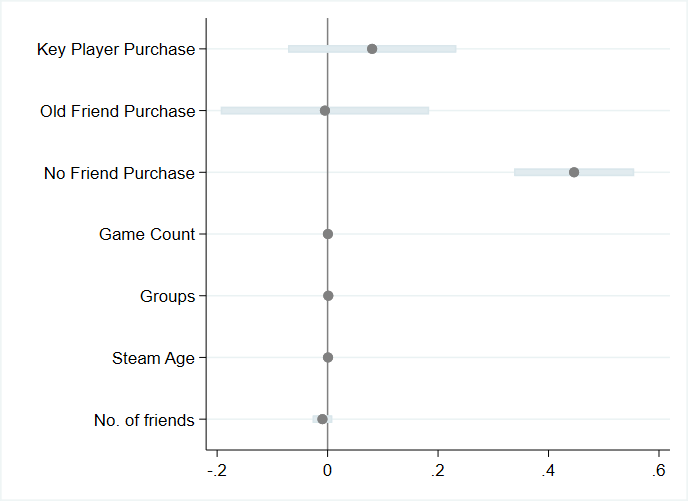}	
    \caption{Peer Effects \& Playtime  - SMB}
\end{subfigure}	
\\
\begin{center}
	\begin{subfigure}[b]{0.45\textwidth}
    \includegraphics[width=7.5cm,height=27cm,keepaspectratio]{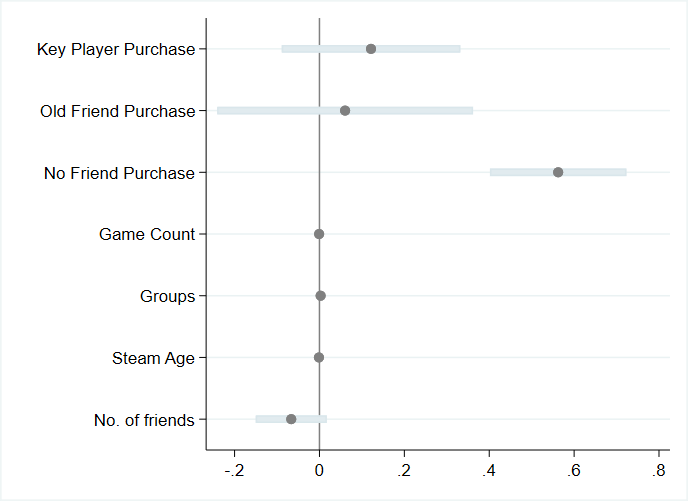}	
    \caption{Peer Effects \& Playtime  - NV}
\end{subfigure}	
\end{center}
\end{figure}
\pagebreak 

\section{Conclusion}  \label{conclusion_1}

Peer effects are an important factor that drives product adoption, but taste projection can also lead to biases in social learning and subsequent enjoyment of the good. This paper presents an empirical analysis of peer effects in the adoption of experiential products, notably video game purchases, and the subsequent engagement of the consumer's with the product. We construct a novel individual-level, high-frequency panel dataset from the world's largest video game platform that includes online networks and video game purchases. Our identification strategy uses plausibly exogenous variation in past adoption decision of second degree friends. We first show that peer effects have a systematic impact on the adoption of computer games. Notably, peer influence is substantially stronger when it stems from long-term friendships (``Old Friends'') rather than from expert or influencer (``Key Player'') connections. This finding aligns with prior research on social influence, highlighting the role of network centrality in shaping consumption choices, particularly for experience goods where quality assessment relies heavily on social cues. We then analyse the subsequent engagement with the games and show that playtime is systematically lower if a user's purchase decision was influenced by peers.  This pattern aligns with the concept of taste projection \citep{gagnonbartsch2024} , where individuals overestimate the preference alignment with others, leading to biased social learning. Players might overestimate their enjoyment of a game chosen by a Key Player, leading to an initial surge in engagement that is not sustained as their actual experience contradicts their projected enjoyment.  

While not directly testing \citet{gagnonbartsch2024} model, the observed playtime patterns offer some first empirical support for the potential of taste projection to shape consumption of experience goods within social networks. This connection underscores the importance of considering cognitive biases in social learning, particularly when examining products where quality is revealed through consumption.

However, as outlined in the paper, there are limitations to using observational data on user engagement with a product to test for taste projection. Further investigation, potentially through experimental studies or the incorporation of richer data on player skill and in-game social interactions, could disentangle these factors and provide a better understanding of the interplay between social influence, individual preferences, and game characteristics in shaping engagement with experience goods.

In summary, our study presents two main findings that might be of interest considering the increasing importance of online networks in consumption decisions: First, there is substantial heterogeneity in peer effects on production adoption depending on the type of online peer. Our results show that peer effects by older friends are stronger than peer effects by key players. Second, peer effects in consumption can potentially lead to biases in consumption decisions and the purchase of goods that lead to less engagement, and maybe less satisfaction, with the good.

\clearpage

\pagebreak

\bibliographystyle{ecca}
\bibliography{first_paper_ref_list.bib}

\begin{thebibliography}{60}
\providecommand{\natexlab}[1]{#1}

\bibitem[{Agarwal \textit{et~al.}(2021)Agarwal, Qian and Zou}]{agarwal2021thy}
\textsc{Agarwal, S.}, \textsc{Qian, W.} and \textsc{Zou, X.} (2021). Thy
  neighbor’s misfortune: Peer effect on consumption. \textit{American
  Economic Journal: Economic Policy}, \textbf{13}~(2), 1--25.

\bibitem[{Aguiar \textit{et~al.}(2021)Aguiar, Bils, Charles and
  Hurst}]{aguiar2021leisure}
\textsc{Aguiar, M.}, \textsc{Bils, M.}, \textsc{Charles, K.~K.} and
  \textsc{Hurst, E.} (2021). Leisure luxuries and the labor supply of young
  men. \textit{Journal of Political Economy}, \textbf{129}~(2), 337--382.

\bibitem[{Ameri \textit{et~al.}(2019)Ameri, Honka and Xie}]{ameri2019word}
\textsc{Ameri, M.}, \textsc{Honka, E.} and \textsc{Xie, Y.} (2019). Word of
  mouth, observed adoptions, and anime-watching decisions: The role of the
  personal vs. the community network. \textit{Marketing Science},
  \textbf{38}~(4), 567--583.

\bibitem[{Bailey \textit{et~al.}(2022)Bailey, Johnston, Kuchler, Stroebel and
  Wong}]{bailey2022peer}
\textsc{Bailey, M.}, \textsc{Johnston, D.}, \textsc{Kuchler, T.},
  \textsc{Stroebel, J.} and \textsc{Wong, A.} (2022). Peer effects in product
  adoption. \textit{American Economic Journal: Applied Economics},
  \textbf{14}~(3), 488--526.

\bibitem[{Banerjee \textit{et~al.}(2013)Banerjee, Chandrasekhar, Duflo and
  Jackson}]{banerjee2013diffusion}
\textsc{Banerjee, A.}, \textsc{Chandrasekhar, A.~G.}, \textsc{Duflo, E.} and
  \textsc{Jackson, M.~O.} (2013). The diffusion of microfinance.
  \textit{Science}, \textbf{341}~(6144), 1236498.

\bibitem[{Barrett \textit{et~al.}(2016)Barrett, Fafchamps, Islam, Malek and
  Pakrashi}]{barrett2016system}
\textsc{Barrett, C.}, \textsc{Fafchamps, M.}, \textsc{Islam, A.},
  \textsc{Malek, A.} and \textsc{Pakrashi, D.} (2016). System of rice
  intensification in rural bangladesh: Adoption, diffusion and impact.
  \textit{Adoption, Diffusion and Impact}.

\bibitem[{Blume \textit{et~al.}(2015)Blume, Brock, Durlauf and
  Jayaraman}]{blume2015linear}
\textsc{Blume, L.~E.}, \textsc{Brock, W.~A.}, \textsc{Durlauf, S.~N.} and
  \textsc{Jayaraman, R.} (2015). Linear social interactions models.
  \textit{Journal of Political Economy}, \textbf{123}~(2), 444--496.

\bibitem[{Bollinger and Gillingham(2012)}]{bollinger2012peer}
\textsc{Bollinger, B.} and \textsc{Gillingham, K.} (2012). Peer effects in the
  diffusion of solar photovoltaic panels. \textit{Marketing Science},
  \textbf{31}~(6), 900--912.

\bibitem[{Bollinger \textit{et~al.}(2022)Bollinger, Gillingham, Kirkpatrick and
  Sexton}]{bollinger2022visibility}
\textsc{---}, \textsc{---}, \textsc{Kirkpatrick, A.~J.} and \textsc{Sexton, S.}
  (2022). Visibility and peer influence in durable good adoption.
  \textit{Marketing Science}, \textbf{41}~(3), 453--476.

\bibitem[{Bramoull{\'e} \textit{et~al.}(2009)Bramoull{\'e}, Djebbari and
  Fortin}]{bramoulle2009identification}
\textsc{Bramoull{\'e}, Y.}, \textsc{Djebbari, H.} and \textsc{Fortin, B.}
  (2009). Identification of peer effects through social networks.
  \textit{Journal of Econometrics}, \textbf{150}~(1), 41--55.

\bibitem[{Bramoullé \textit{et~al.}(2020)Bramoullé, Djebbari and
  Fortin}]{bramoulle2020}
\textsc{Bramoullé, Y.}, \textsc{Djebbari, H.} and \textsc{Fortin, B.} (2020).
  Peer effects in networks: A survey. \textit{Annual Review of Economics},
  \textbf{12}, 603--629.

\bibitem[{Breza \textit{et~al.}(2020)Breza, Chandrasekhar, McCormick and
  Pan}]{breza2020using}
\textsc{Breza, E.}, \textsc{Chandrasekhar, A.~G.}, \textsc{McCormick, T.~H.}
  and \textsc{Pan, M.} (2020). Using aggregated relational data to feasibly
  identify network structure without network data. \textit{American Economic
  Review}, \textbf{110}~(8), 2454--2484.

\bibitem[{Burke and Sass(2013)}]{burke2013classroom}
\textsc{Burke, M.~A.} and \textsc{Sass, T.~R.} (2013). Classroom peer effects
  and student achievement. \textit{Journal of Labor Economics},
  \textbf{31}~(1), 51--82.

\bibitem[{Calv{\'o}-Armengol \textit{et~al.}(2009)Calv{\'o}-Armengol,
  Patacchini and Zenou}]{calvo2009peer}
\textsc{Calv{\'o}-Armengol, A.}, \textsc{Patacchini, E.} and \textsc{Zenou, Y.}
  (2009). Peer effects and social networks in education. \textit{Review of
  Economic Studies}, \textbf{76}~(4).

\bibitem[{Campbell \textit{et~al.}(2020)Campbell, Leister and
  Zenou}]{campbell2020word}
\textsc{Campbell, A.}, \textsc{Leister, C.~M.} and \textsc{Zenou, Y.} (2020).
  Word-of-mouth communication and search. \textit{RAND Journal of Economics},
  \textbf{51}~(3), 676--712.

\bibitem[{Comola and Prina(2021)}]{comola2021treatment}
\textsc{Comola, M.} and \textsc{Prina, S.} (2021). Treatment effect accounting
  for network changes. \textit{Review of Economics and Statistics},
  \textbf{103}~(3), 597--604.

\bibitem[{Conley and Udry(2010)}]{conley2010learning}
\textsc{Conley, T.~G.} and \textsc{Udry, C.~R.} (2010). Learning about a new
  technology: Pineapple in ghana. \textit{American Economic Review},
  \textbf{100}~(1), 35--69.

\bibitem[{Cunningham \textit{et~al.}(2016)Cunningham, Engelst{\"a}tter and
  Ward}]{cunningham2016violent}
\textsc{Cunningham, S.}, \textsc{Engelst{\"a}tter, B.} and \textsc{Ward, M.~R.}
  (2016). Violent video games and violent crime. \textit{Southern Economic
  Journal}, \textbf{82}~(4), 1247--1265.

\bibitem[{Dautovic(2023)}]{dautovic_2021}
\textsc{Dautovic, G.} (2023). The rise of the virtual empire: Video game
  industry statistics for 2024.

\bibitem[{De~Giorgi \textit{et~al.}(2020)De~Giorgi, Frederiksen and
  Pistaferri}]{de2020consumption}
\textsc{De~Giorgi, G.}, \textsc{Frederiksen, A.} and \textsc{Pistaferri, L.}
  (2020). Consumption network effects. \textit{Review of Economic Studies},
  \textbf{87}~(1), 130--163.

\bibitem[{De~Giorgi \textit{et~al.}(2010)De~Giorgi, Pellizzari and
  Redaelli}]{de2010identification}
\textsc{---}, \textsc{Pellizzari, M.} and \textsc{Redaelli, S.} (2010).
  Identification of social interactions through partially overlapping peer
  groups. \textit{American Economic Journal: Applied Economics},
  \textbf{2}~(2), 241--275.

\bibitem[{de~Paula \textit{et~al.}(2024)de~Paula, Rasul and
  Souza}]{depaula2024_panelsocialties}
\textsc{de~Paula, A.}, \textsc{Rasul, I.} and \textsc{Souza, P.~C.} (2024).
  {Identifying network ties from panel data: Theory and an application to tax
  competition}. \textit{Review of Economic Studies}, p. Advance online
  publication.

\bibitem[{Engelmann and Strobel(2012)}]{engelmann2012deconstruction}
\textsc{Engelmann, D.} and \textsc{Strobel, M.} (2012). Deconstruction and
  reconstruction of an anomaly. \textit{Games and Economic Behavior},
  \textbf{76}~(2), 678--689.

\bibitem[{Foster and Rosenzweig(1995)}]{foster1995learning}
\textsc{Foster, A.~D.} and \textsc{Rosenzweig, M.~R.} (1995). Learning by doing
  and learning from others: Human capital and technical change in agriculture.
  \textit{Journal of Political Economy}, \textbf{103}~(6), 1176--1209.

\bibitem[{Gagnon-Bartsch and Rosato(2024)}]{gagnonbartsch2024}
\textsc{Gagnon-Bartsch, T.} and \textsc{Rosato, A.} (2024). Quality is in the
  eye of the beholder: Taste projection in markets with observational learning.
  \textit{American Economic Review}, advance online publication.

\bibitem[{{Game Developers Conference}(2019)}]{gdc2019}
\textsc{{Game Developers Conference}} (2019). State of the game industry 2019.

\bibitem[{Gandhi \textit{et~al.}(2024)Gandhi, Giuliano, Guan, Keefer, McDonald,
  Pagel and Tasoff}]{Gandhi2025}
\textsc{Gandhi, A.}, \textsc{Giuliano, P.}, \textsc{Guan, E.}, \textsc{Keefer,
  Q.}, \textsc{McDonald, C.}, \textsc{Pagel, M.} and \textsc{Tasoff, J.}
  (2024). \textit{Beliefs that Entertain}. Working Paper 32295, National Bureau
  of Economic Research.

\bibitem[{Granic \textit{et~al.}(2014)Granic, Lobel and
  Engels}]{granic2014benefits}
\textsc{Granic, I.}, \textsc{Lobel, A.} and \textsc{Engels, R.~C.} (2014). The
  benefits of playing video games. \textit{American psychologist},
  \textbf{69}~(1), 66.

\bibitem[{Hanushek \textit{et~al.}(2003)Hanushek, Kain, Markman and
  Rivkin}]{hanushek2003does}
\textsc{Hanushek, E.~A.}, \textsc{Kain, J.~F.}, \textsc{Markman, J.~M.} and
  \textsc{Rivkin, S.~G.} (2003). Does peer ability affect student achievement?
  \textit{Journal of Applied Econometrics}, \textbf{18}~(5), 527--544.

\bibitem[{Hoffman(2017)}]{hoffman2017}
\textsc{Hoffman, R.} (2017). \textit{Following the Peers: The Role of Social
  Networks for Health Care Utilization in the {P}hilippines}. Working Paper~17,
  Vienna Institute of Demography (VID).

\bibitem[{Humphries(2018)}]{humphries_pcmag_2018}
\textsc{Humphries, M.} (2018). Steam accounts just got a lot more private.

\bibitem[{Jochmans(2023)}]{jochmans2023peer}
\textsc{Jochmans, K.} (2023). Peer effects and endogenous social interactions.
  \textit{Journal of Econometrics}, \textbf{235}~(2), 1203--1214.

\bibitem[{Katona \textit{et~al.}(2011)Katona, Zubcsek and
  Sarvary}]{katona2011network}
\textsc{Katona, Z.}, \textsc{Zubcsek, P.~P.} and \textsc{Sarvary, M.} (2011).
  Network effects and personal influences: The diffusion of an online social
  network. \textit{Journal of Marketing Research}, \textbf{48}~(3), 425--443.

\bibitem[{Li \textit{et~al.}(2022)Li, He, Liu and Ping}]{li2022peer}
\textsc{Li, Y.}, \textsc{He, J.}, \textsc{Liu, C.} and \textsc{Ping, Y.}
  (2022). Peer influence in the adoption of video games. \textit{International
  Journal of E-Business Research (IJEBR)}, \textbf{18}~(1), 1--16.

\bibitem[{Ma \textit{et~al.}(2015)Ma, Krishnan and Montgomery}]{ma2015latent}
\textsc{Ma, L.}, \textsc{Krishnan, R.} and \textsc{Montgomery, A.~L.} (2015).
  Latent homophily or social influence? an empirical analysis of purchase
  within a social network. \textit{Management Science}, \textbf{61}~(2),
  454--473.

\bibitem[{Manski(1993)}]{manski1993identification}
\textsc{Manski, C.~F.} (1993). Identification of endogenous social effects: The
  reflection problem. \textit{Review of Economic Studies}, \textbf{60}~(3),
  531--542.

\bibitem[{Mas and Moretti(2009)}]{mas2009peers}
\textsc{Mas, A.} and \textsc{Moretti, E.} (2009). Peers at work.
  \textit{American Economic Review}, \textbf{99}~(1), 112--145.

\bibitem[{Mc{E}lroy(2013)}]{mcelroy_polygon_2013}
\textsc{Mc{E}lroy, G.} (2013). How an indie game made a word-of-mouth rise to
  the top of the e{S}hop charts.

\bibitem[{McPherson \textit{et~al.}(2001)McPherson, Smith-Lovin and
  Cook}]{mcpherson2001birds}
\textsc{McPherson, M.}, \textsc{Smith-Lovin, L.} and \textsc{Cook, J.~M.}
  (2001). Birds of a feather: Homophily in social networks. \textit{Annual
  review of sociology}, \textbf{27}~(1), 415--444.

\bibitem[{Miller and Mobarak(2015)}]{miller2015learning}
\textsc{Miller, G.} and \textsc{Mobarak, A.~M.} (2015). Learning about new
  technologies through social networks: experimental evidence on nontraditional
  stoves in bangladesh. \textit{Marketing Science}, \textbf{34}~(4), 480--499.

\bibitem[{{Mordor Intelligence}(2021)}]{mordorintelligence}
\textsc{{Mordor Intelligence}} (2021). Gaming industry size \& share analysis -
  growth trends \& forecasts (2021--2029).

\bibitem[{Moretti(2011)}]{moretti2011social}
\textsc{Moretti, E.} (2011). Social learning and peer effects in consumption:
  Evidence from movie sales. \textit{Review of Economic Studies},
  \textbf{78}~(1), 356--393.

\bibitem[{Munshi(2003)}]{munshi2003networks}
\textsc{Munshi, K.} (2003). Networks in the modern economy: Mexican migrants in
  the us labor market. \textit{Quarterly Journal of Economics},
  \textbf{118}~(2), 549--599.

\bibitem[{Nair \textit{et~al.}(2010)Nair, Manchanda and
  Bhatia}]{nair2010asymmetric}
\textsc{Nair, H.~S.}, \textsc{Manchanda, P.} and \textsc{Bhatia, T.} (2010).
  Asymmetric social interactions in physician prescription behavior: The role
  of opinion leaders. \textit{Journal of Marketing Research}, \textbf{47}~(5),
  883--895.

\bibitem[{Nauze(2023)}]{lanauze2021}
\textsc{Nauze, A.~L.} (2023). {Motivation Crowding in Peer Effects: The Effect
  of Solar Subsidies on Green Power Purchases}. \textit{Review of Economics and
  Statistics}, \textbf{105}~(6), 1465--1480.

\bibitem[{O'Neill \textit{et~al.}(2016)O'Neill, Vaziripour, Wu and
  Zappala}]{ONeilletal2016}
\textsc{O'Neill, M.}, \textsc{Vaziripour, E.}, \textsc{Wu, J.} and
  \textsc{Zappala, D.} (2016). Condensing {S}team: Distilling the diversity of
  gamer behavior. In \textit{Proceedings of the 2016 Internet Measurement
  Conference}, IMC '16, New York, NY, USA: Association for Computing Machinery,
  p. 81–95.

\bibitem[{Orhun and Urminsky(2013)}]{orhun2013conditional}
\textsc{Orhun, A.~Y.} and \textsc{Urminsky, O.} (2013). Conditional projection:
  How own evaluations influence beliefs about others whose choices are known.
  \textit{Journal of Marketing Research}, \textbf{50}~(1), 111--124.

\bibitem[{Patacchini \textit{et~al.}(2017)Patacchini, Rainone and
  Zenou}]{patacchini2017heterog}
\textsc{Patacchini, E.}, \textsc{Rainone, E.} and \textsc{Zenou, Y.} (2017).
  Heterogeneous peer effects in education. \textit{Journal of Economic Behavior
  {\&} Organization}, \textbf{134}, 190--227.

\bibitem[{Patacchini and Zenou(2008)}]{patacchini2008strength}
\textsc{---} and \textsc{Zenou, Y.} (2008). The strength of weak ties in crime.
  \textit{European Economic Review}, \textbf{52}~(2), 209--236.

\bibitem[{Robalino and Macy(2018)}]{robalino2018peer}
\textsc{Robalino, J.~D.} and \textsc{Macy, M.} (2018). Peer effects on
  adolescent smoking: Are popular teens more influential? \textit{PloS one},
  \textbf{13}~(7), e0189360.

\bibitem[{Ross \textit{et~al.}(1977)Ross, Greene and House}]{ross1977false}
\textsc{Ross, L.}, \textsc{Greene, D.} and \textsc{House, P.} (1977). The
  “false consensus effect”: An egocentric bias in social perception and
  attribution processes. \textit{Journal of Experimental Social Psychology},
  \textbf{13}~(3), 279--301.

\bibitem[{Roth(2015)}]{roth2015conspicuous}
\textsc{Roth, C.} (2015). Conspicuous consumption and peer effects: evidence
  from a randomized field experiment. \textit{Available at SSRN 2586716}.

\bibitem[{Ryan(2017)}]{ryan2017measurement}
\textsc{Ryan, C.} (2017). Measurement of peer effects. \textit{Australian
  Economic Review}, \textbf{50}~(1), 121--129.

\bibitem[{Shakya \textit{et~al.}(2015)Shakya, Christakis and
  Fowler}]{shakya2015social}
\textsc{Shakya, H.~B.}, \textsc{Christakis, N.~A.} and \textsc{Fowler, J.~H.}
  (2015). Social network predictors of latrine ownership. \textit{Social
  Science {\&} Medicine}, \textbf{125}, 129--138.

\bibitem[{Siciliano(2016)}]{siciliano2016s}
\textsc{Siciliano, M.~D.} (2016). It’s the quality not the quantity of ties
  that matters: Social networks and self-efficacy beliefs. \textit{American
  Educational Research Journal}, \textbf{53}~(2), 227--262.

\bibitem[{Steam(2021)}]{steam2021review}
\textsc{Steam} (2021). Steam - 2021 year in review.

\bibitem[{Tud{\'o}n(2022)}]{tudon2022distilling}
\textsc{Tud{\'o}n, J.} (2022). Distilling network effects from steam.
  \textit{Quantitative Marketing and Economics}, \textbf{20}~(3), 293--312.

\bibitem[{{Valve Developer Community}(2023)}]{valvedev_steamwebapi_page}
\textsc{{Valve Developer Community}} (2023). {Steam Web API}.

\bibitem[{Ward(2011)}]{ward2011video}
\textsc{Ward, M.~R.} (2011). Video games and crime. \textit{Contemporary
  economic policy}, \textbf{29}~(2), 261--273.

\bibitem[{Zenou \textit{et~al.}(2021)Zenou, Islam, Vlassopoulos and
  Zhang}]{zenouislam_repec}
\textsc{Zenou, Y.}, \textsc{Islam, A.}, \textsc{Vlassopoulos, M.} and
  \textsc{Zhang, X.} (2021). \textit{{Centrality-Based Spillover Effects}}.
  CEPR Discussion Papers 16321, C.E.P.R. Discussion Papers.

\end{thebibliography}

\clearpage


\clearpage


\appendix

\section{The Steam Platform - Additional Information}  \label{appendix_sect_steamplatform}

\renewcommand{\thefigure}{A\arabic{figure}}
\renewcommand{\thetable}{A\arabic{table}}
\setcounter{figure}{0} 
\setcounter{table}{0}

\begin{figure}[!h]%
\centering 
\caption{Steam Platform - Store Front}
    \includegraphics[width=12cm]{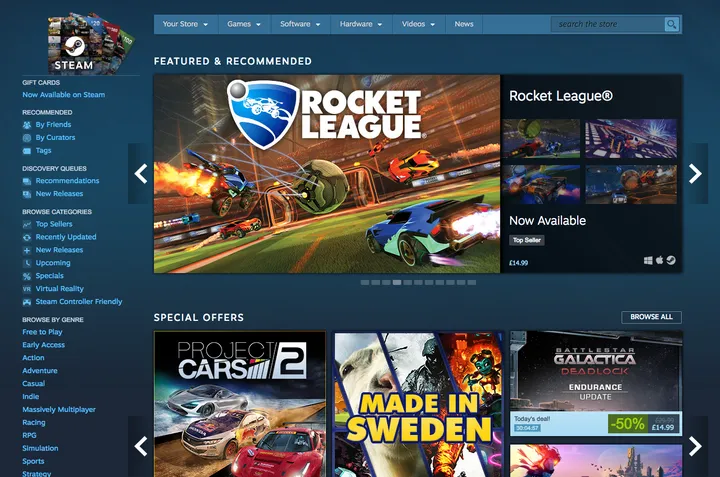} %
    \label{appfig:steamstore}%
\end{figure}

Figure \ref{appfig:steamstore} presents a screenshot of the steam store.

\begin{figure}[!h]%
\centering 
\caption{Steam Platform - Store Front}
    \includegraphics[width=12cm]{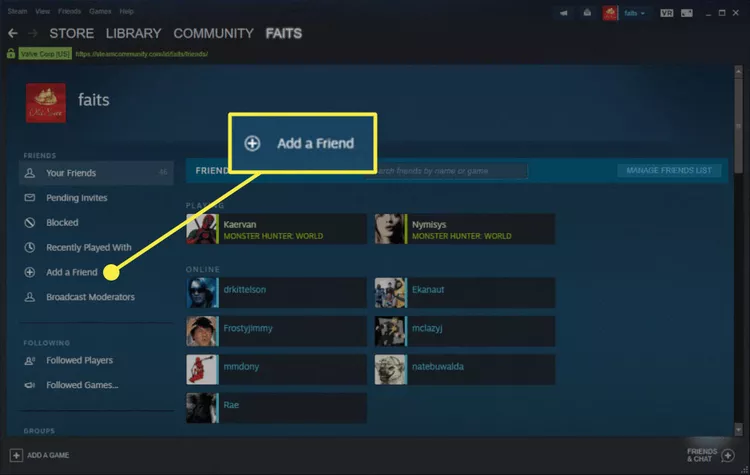} %
    \label{appfig:steamfriends}%
\end{figure}

Figure \ref{appfig:steamfriends} presents a screenshot of a users profile page.  The menu on the left has a function allowing the user to add new friends to their friends list. The center panel shows the user which friends a currently online and for friends who are currently engaged in a game it also shows the title of the game (in yellow) underneath the friend's username. 

\begin{figure}[!h]%
\centering 
\caption{Steam Platform - Example of Public Game Achievement Data}
    \includegraphics[width=12cm]{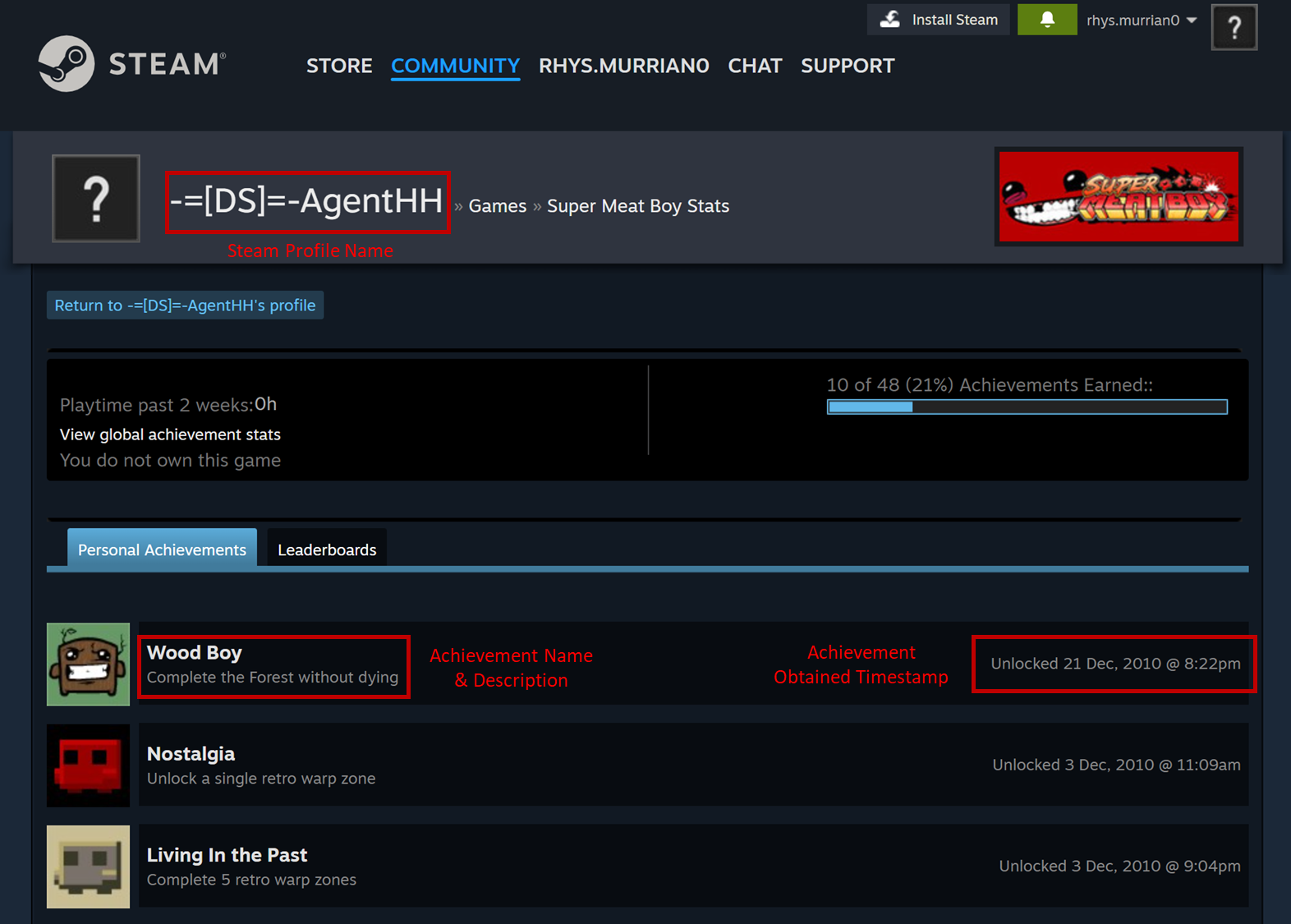} %
    \label{appfig:steamachieve}%
\end{figure}

Finally, Figure \ref{appfig:steamachieve} is a screenshot from a player's (\textit{-=[DS]=-AgentHH}) public Steam community profile page that displays achievements for the game \textit{Super Meat Boy}. It includes various sections of information. Areas highlighted in red are the parts of the profile that have been scraped to collect the achievement data that was subsequently used to determine the date the player purchased Super Meat Boy. The Steam profile name, highlighted as Steam Profile Name, identifies the player whose achievements are being shown which is used to link the achievement data to our main dataset. The game Super Meat Boy is also indicated at the top right, specifying the game for which these achievements have been unlocked. Below the profile name, the image shows that the player has earned 10 out of 48 possible achievements, which represents 21\% of the total. The lower part of the image displays individual achievements, each accompanied by details such as the achievement name and description, highlighted for the achievement ``Wood Boy''.  Additionally, the date and time when this achievement was unlocked, marked as ``Unlocked 21 Dec, 2010 @ 8:22pm,'' are also highlighted. These highlighted elements—Steam profile name, achievement details, and timestamps—are critical for collecting data on player achievements. The profile name links the achievements to a specific player, the achievement details provide insight into the milestones or challenges completed, and the timestamps track when these achievements were unlocked. We then use the timestamp of the player's first achievement as the proxy for the purchase date.

\clearpage

\section {Data Appendix} \label{appendix_sect_data}

\renewcommand{\thefigure}{B\arabic{figure}}
\renewcommand{\thetable}{B\arabic{table}}
\setcounter{figure}{0} 
\setcounter{table}{0}

\subsection{Steam Friendship Network}
\begin{figure}[htbp]
    \centering
    \caption{Steam Friendships Network Data - Illustrative Example}
    \begin{subfigure}[b]{0.5\textwidth}  
        \centering
        \includegraphics[width=\textwidth]{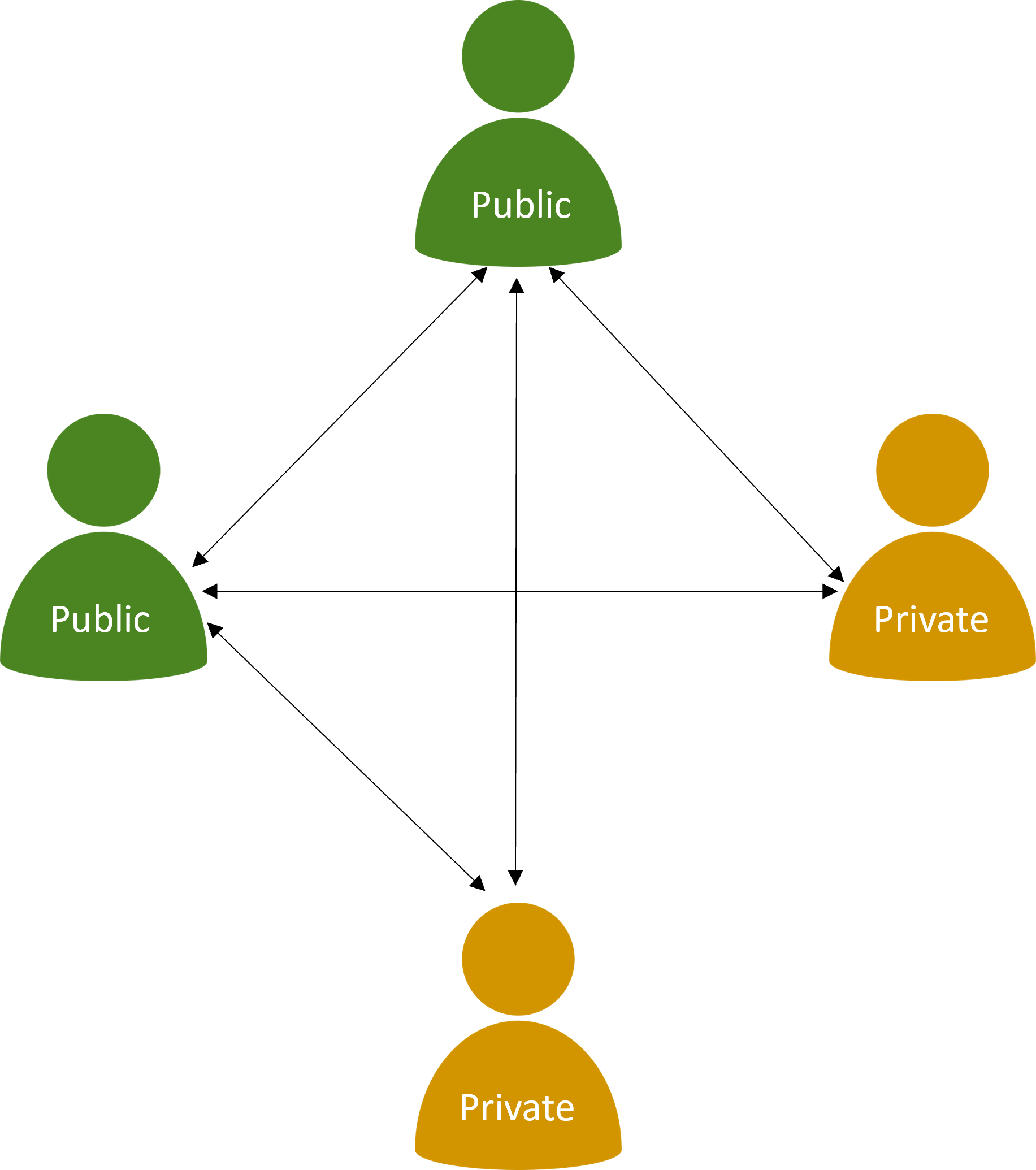}
    \end{subfigure}
    \label{fig:steam_friendships_diagram}

    \vspace{6pt} 

    \begin{minipage}{\textwidth}
        \footnotesize
        \textit{Note:} The figure above presents an illustrative example of the Steam friendship network collected in the \citet{ONeilletal2016} dataset. Friendships between two private Steam users are unobservable from the data. However, as 93.7\% of Steam users in the dataset were public users at the time the data was collated, the number of private-private Steam friendships is assumed to comprise a minimal proportion of the total number of Steam friendships.
\end{minipage}

\end{figure}

The network of friends on Steam collated by \citet{ONeilletal2016} was obtained via the Steam Web API. However, the friendship data could only be ascertained for public users. Hence, the constructed online network only contains (bi-directional) links for public-public and public-private users, but not private-private users. This implies that the constructed online network is incomplete. The incompleteness of the network is illustrated by Figure \ref{fig:steam_friendships_diagram}. In the example figure, there are four Steam users, two of which have public accounts and two of which have private accounts. All four users are friends with each other. While we can observe the links between the public-public and public-private users, we cannot observe the link between the private-private users, even though they are friends. As such, it cannot be determined whether a link exists between two private users. However, this limitation is likely to be minor as public users account for 93.7\% of all users in the dataset.



An additional limitation of the network is the general limit of 250 friends for each user. However, the limit can be artificially increased through a user linking their Steam account to their Facebook account to increase their friendship limit by an additional 50 users or by increasing their friendship limit by 5 friends every time they obtain a new 'Steam' level.\footnote{To keep reaching new \emph{Steam Levels}, a user must earn a certain amount of experience points (XP). XP is gained by purchasing more games, collecting trading cards/crafting badges for games, or participating in Steam-based community events.} This limitation is minor as the majority of Steam users do not reach the 250 friendship limit.\footnote{0.01\% of users in the  \citet{ONeilletal2016} dataset have reached this number of friendships.}


\subsection{Purchase Dates}
Given the unavailability of data regarding the purchase dates of games within each Steam user's library, we make the strong assumption that the timestamp of when a user obtained their first achievement in SMB (or NV) is a suitable proxy for when that user purchased the game. This assumption rests on two underlying conditions: One, that a user who purchases a game plays it almost immediately after purchasing it; and two, that the time taken to obtain an achievement in the game is relatively short. While we cannot directly test the first condition, it appears reasonable - Video games are a leisure good, and hence, will likely be purchased only when a user has the time to play it. With regard to the second condition, we provide empirical evidence to further support the claim. Table \ref{table:smb_one_achievement_stats} presents the summary statistics for the playtime of SMB players who have only obtained one achievement within the \citet{ONeilletal2016} dataset (for whom player achievement data was available). The mean playtime for SMB players who obtained only one achievement is 133 minutes, implying that the average player takes slightly over 2 hours to obtain their first achievement in SMB. However, this value is likely to be skewed by outliers as illustrated by the maximum playtime of 13,434 minutes (over 9 days) in the sample. A more appropriate indicator would be the median, which shows that the 50th percentile of SMB players in the sample obtain an achievement within 73 minutes of playing SMB. Given that the panel data is aggregated up to a weekly level, it is reasonable to expect that a player would obtain an achievement within the same week as the purchase date.

\begin{table}[htbp]

\caption{Summary Statistics - Super Meat Boy Players with One Achievement} \label{table:smb_one_achievement_stats}

\begin{adjustbox}{width=1\textwidth}
\fontsize{10}{12}\selectfont

\begin{tabular}{lrrrrrr}
\toprule
\toprule
\textbf{Variable} &     \textbf{Observations} &   \textbf{Mean} &  \textbf{Std. dev.}  &  \textbf{Min} &   \textbf{Median} &   \textbf{Max} \\
\midrule
SMB Playtime        &  48,605 &  133.29 &   233.52 &  1  &  73 &   13,434 \\
\bottomrule
\bottomrule
\end{tabular}
\end{adjustbox}

\vspace{6pt}

\begin{minipage}{\textwidth}
        \footnotesize
        \textit{Notes:} This table presents summary statistics for the sample of players who have obtained only one achievement in SMB in the  \citet{ONeilletal2016} dataset (for whom player achievement data was available). Values of SMB playtime are presented in minutes.
\end{minipage}

\end{table}

\clearpage

\section{Additional Tables}   \label{appendix_sect_additional_tables}

\renewcommand{\thefigure}{C\arabic{figure}}
\renewcommand{\thetable}{C\arabic{table}}
\setcounter{figure}{0} 
\setcounter{table}{0}



\begin{table}[h!]

\centering

\caption{\citet{ONeilletal2016} dataset - Descriptions}  \label{table:oneill_dataset_desc}
\begin{adjustbox}{}  
{
\setlength{\tabcolsep}{10pt}  
\def\sym#1{\ifmmode^{#1}\else\(^{#1}\)\fi}
\begin{tabular}{p{0.275\linewidth} p{0.675\linewidth}}  
\toprule
\toprule
\textbf{Table} & \textbf{Description}  \\
\midrule
Achievement\_Percentages	&	For each achievement within each Steam app, the percentage of users who have completed the achievement.	\\
App\_ID\_Info	&	Information provided for each Steam App.	\\
Friends	&	Friend network for each Steam user.  Note, a user’s friendship network is only available when their profile is set to ‘Public’ within the \textit{Player\_Summaries} table. 	\\
Games\_1	&	Games library for each Steam user from the first web crawl.  Note, a user’s games library is only available when their profile is set to ‘Public’ within the \textit{Player\_Summaries} table.	\\
Games\_2	&	Games library for each Steam user from the second web crawl.  Note, a user’s games library is only available when their profile is set to ‘Public’ within the \textit{Player\_Summaries} table.	\\
Games\_Daily	&	Playtime for each user for each Steam app over the last two weeks at the time of data collection. Note, this information is only available when their profile is set to ‘Public’ within the \textit{Player\_Summaries} table.	\\
Games\_Developers	&	Developer information for each Steam app.	\\
Games\_Genres	&	Genre for each Steam app.	\\
Games\_Publishers	&	Publisher information for each Steam app.	\\
Groups	&	Provides the list of groups/clans that each Steam user belongs to.  Note, the groups that a user has joined is only available when their profile is set to ‘Public’ within the \textit{Player\_Summaries} table.	\\
Player\_Summaries	&	Provides summary information for each Steam user.	\\
\bottomrule
\bottomrule
\end{tabular}
}
\end{adjustbox}

\vspace{6pt}
\begin{minipage}{\linewidth}
\footnotesize\textit{Notes:} The table above lists and describes all of the tables that comprise the \citet{ONeilletal2016} dataset.
\end{minipage}

\end{table}

\newpage


\begin{table}[!h]
\centering
\caption{Summary Statistics - Peer Effects and Playtime}
\label{table:playtimedes}
\begin{tabular}{@{\extracolsep{5pt}}lcccccc}

\\[-1.8ex]\hline \hline \\[-1.8ex]
\textbf{Variable}	&	\textbf{Mean}	&	\textbf{Std. dev.}	&	\textbf{Min}	&	\textbf{Max}	\\\hline

Playtime	&	27.921	&	55.110	&	1	&	728	\\
KP Friend Purchase&	0.163	&	0.369	&	0	&	1	\\
Old Friend Purchase&	0.083	&	0.275	&	0	&	1	\\
\# of Games&	122.269	&	94.015	&	1	&	1011	\\
\# of Groups&	13.568	&	30.677	&	0	&	642	\\
Start week&	62.723	&	66.803	&	0	&	224	\\
Friends &	0.457	&	1.525	&	0	&	36	\\
New Vegas Game&	0.469	&	0.499	&	0	&	1	\\
Super Meat Boy Game &	0.537	&	0.499	&	0	&	1	\\

\hline \hline \\ 
\end{tabular}

\vspace{6pt}
\begin{minipage}{\linewidth}
\footnotesize\textit{Notes:} This table presents the descriptive statistics for the data used in Section \ref{sec:playtime} Peer Effects and Playtime. No. of obs.: 3,622
\end{minipage}
\end{table}

\pagebreak

\begin{table}[H]

\caption{Peer Effects and Playtime \label{tab:pt}}
\centering
 
\resizebox{0.9\textwidth}{!}{
\begin{tabular}[t]{lllcccc}
\hline \hline
 &  &  & \multicolumn{4}{c}{Dependent Variable: Ln(Playtime)}\\
 &  &  & (1) & (2) & (3) & (4)  \\
\midrule
\addlinespace[0.3em]

\addlinespace[0.3em]
\multicolumn{6}{l}{\textbf{A: SMB and NV Combined}}\\


\hspace{1em} & &Key Player Purchase & -0.0806 &  &  & 0.1070 \\
\hspace{1em} & && (0.0647) &  &  & (0.0664) \\
\hspace{1em}& &Old Friend Purchase &  & -0.1637** &  & 0.0128 \\
\hspace{1em}& & &  & (0.0831) &  & (0.0848) \\
\hspace{1em}& &No Friend Purchase &  &  & 0.4822*** & 0.5062*** \\
\hspace{1em}& & &  &  & (0.0469) & (0.0494) \\
  \cmidrule{2-7}
\hspace{1em}& &Observations & 3,496 & 3,496 & 3,496 & 3,496 \\

 \cmidrule{1-7}
\addlinespace[0.3em]
\multicolumn{6}{l}{\textbf{B: SMB Only}}\\
 
\hspace{1em} & &Key Player Purchase & -0.0655 &  &  & 0.0807 \\
\hspace{1em} & & & (0.0749) &  &  & (0.0777) \\
\hspace{1em} & &Old Friend Purchase &  & -0.1634* &  & -0.0048 \\
\hspace{1em} & & &  & (0.0949) &  & (0.0961) \\
\hspace{1em} & &No Friend Purchase &  &  & 0.4321*** & 0.4463*** \\
\hspace{1em} & & &  &  & (0.0528) & (0.0554) \\
  \cmidrule{2-7}
\hspace{1em}& &Observations& 1,819 & 1,819 & 1,819 & 1,819 \\

 \cmidrule{1-7}
\addlinespace[0.3em]
\multicolumn{6}{l}{\textbf{C: NV Only}}\\

\hspace{1em} & &Key Player Purchase & -0.1099 &  &  & 0.1215 \\
\hspace{1em} & & & (0.1040) &  &  & (0.1073) \\
\hspace{1em} & &Old Friend Purchase &  & -0.1337 &  & 0.0604 \\
\hspace{1em} & & &  & (0.1508) &  & (0.1537) \\
\hspace{1em} & &No Friend Purchase &  &  & 0.5285*** & 0.5625*** \\
\hspace{1em} & & &  &  & (0.0775) & (0.0820) \\
  \cmidrule{2-7}
\hspace{1em}& &Observations& 1,697 & 1,697 & 1,697 & 1,697 \\

\hline \hline
\multicolumn{5}{p{0.8\textwidth}}{\scriptsize{\textit{Notes:} This table presents the raw correlations between various forms of radio indicators at the county level. Robust standard errors in parentheses.*** p$<$0.01, ** p$<$0.05, * p$<$0.1}} \\
\end{tabular}
}

 
\end{table}

\end{document}